# A Simple and Efficient Lattice Summation Method for Metallic Electrodes in Constant Potential Molecular Dynamics Simulation

Haoyu LI[1], Peiyao WANG[3], Jefferson Zhe LIU[4,a)], Gengping JIANG[1,2,b)]

[1] State Key Laboratory of Refractories and Metallurgy, Hubei Province Key Laboratory of Systems Science in Metallurgical Process, International Research Institute for Steel Technology, Collaborative Innovation Center for Advanced Steels, Wuhan University of Science and Technology, Wuhan, 430081, China

[2] Department of Applied Physics, College of Science, Wuhan University of Science and Technology, Wuhan, 430081, China

[3] Department of Chemical Engineering, The University of Melbourne, Parkville, Victoria, 3010, Australia

[4] Department of Mechanical Engineering, The University of Melbourne, Parkville, Victoria, 3010, Australia.

[a,b)] Author to whom correspondence should be addressed: zhe.liu@unimelb.edu.au, Gengpingjiang@wust.edu.cn

## Abstract

The constant potential molecular dynamics simulation method proposed by Siepmann and Sprik and reformulated later by Reed (SR-CPM) has been widely employed to investigate the metallic electrolyte/electrode interfaces, especially for conducting nanochannels with complex connectivity, *e.g.*, carbide-derived carbon or graphene-assembled membrane. This work makes substantial extensions of this seminal SR-CPM approach. First, we introduce two numerical techniques to determine electrode atom charges with an order of magnitude improvement in computational efficiency compared with those widely employed methods. The first numerical technique dramatically accelerates the calculation of the Ewald interaction matrix $\mathbf{E}$, which takes advantage of the existing highly optimised electrostatic codes. The second technique introduces a new preconditioning technique in the conjugate gradient method to considerably increase the computational efficiency of a linear equation system that determines electrode atomic charges. Our improved SR-CPM implemented in the LAMMPS package can handle extra-large systems, *e.g.*, over 8.1 million electrode atoms. Second, after demonstrating the importance of the electroneutrality constraint, we propose a two-step method to enforce electroneutrality in the following post-treatment step, applicable for matrix and iterative techniques. Third, we propose a solid theoretical analysis for the adjustable parameter $\alpha_i$ (namely the atomic Hubbard-U $U_i^0$), which is arbitrarily selected in many SR-CPM simulation practices. We proposed that the optimised $\alpha_i$ or $U_i^0$ should compensate for the electrical potential/energy discrepancy between the discrete atomistic model and the continuum limit. The analytical and optimal $\alpha_i^0$ values are derived for a series of 2D materials.

## I. Introduction

The electric double layers (EDLs) at solid/electrolyte interfaces play a fundamental role in various disciplines, from electrochemistry, colloidal chemistry, pharmaceuticals to molecular biology[1]. To understand the mechanisms underlying experimental observations, atomistic simulations such as molecular dynamics (MD) or Monte Carlo (MC) are extensively adopted to explore the microscopic structure of EDLs. At present, most works that studied the EDLs at the electrified metallic surface used the Fixed Charge Method (FCM), in which the atomic charges of electrode atoms are preset at fixed values, and the electrostatic interaction between the electrode atoms and electrolyte molecules/ions is calculated via the Ewald summation. Nevertheless, experimental and theoretical studies have indicated that the charge fluctuation or polarization of the electrode is critical for the interface structural and dynamic properties.[2-4] The polarizable model distinctively changes the characteristics of the electrolyte and produces many unexpected phenomena at the electrified surface or in the nanochannel. For example, the polarizable metallic nanochannel would considerably enhance ion adsorption. The resultant super-ionic state is proposed via a Monte Carlo simulation[5], studied by the following Molecular Dynamics (MD)[6] and verified by the NMR experiment[7]. The metal surface is also proposed to accelerate ion transport inside the nanochannel.[8] The voltage (potential) control is compatible with the experimental electrochemical working conditions instead of the FCM.[9] Therefore, constant potential methods (CPMs) have been proposed to simulate the natural charge fluctuation on the conducting metal electrode. It allows for a precise evaluation of the polarisation effect on the atomic scale.

The CPM approach introduced by Siepmann and Sprik[10] and reformulated later by Reed et al.[11], referred to as SR-CPM, is the most extensively used CPM method in atomistic simulations. It allows the fluctuation and redistribution of the electrode atomic charges $q_i$ to mimic the metallic surface under a constant electric potential constraint $\chi_i$. Compared with other CPMs, which often have limitations on electrode geometry (e.g., regular geometry[12-16] or a closed topological surface[17, 18]), SR-CPM can handle any arbitrary electrode geometry. It is, therefore, feasible to study the actual electrode materials with the complex nanochannel networks, for example, the promising 2D graphene membrane with a large number of pin-holes and dangling bonds[19, 20] and 3D nanoporous carbide-derived carbon.[21, 22] In addition, the direct control of the electrode potential in SR-CPM simulations provides a method to derive a continuous differential capacitance over a range of potentials.[23] Thus, significant attention has been devoted to further developing and applying SR-CPM techniques. For example, Gingrich and Wilson corrected some derivations for the energy functional of the SR-CPM technique under the 3D or 2D periodic conditions[24] and suggested that the Gaussian charge interaction in SR-CPM studies could be handled using the conventional point charge lattice summation technique. Vatamanua et al. simplified the calculation of the SR-CPM by removing the Gaussian screening.[25] Liang et al. presented a similar electrochemical MD simulation using the third generation of the charge-optimized many-body (COMB3) potential, in which the S-type slater charges and the Wolf summation were employed as the electrode charges and the evaluation method, respectively.[26] A series of substantial improvements and developments has been made by Scalfi and her collaborators, including the crucial role of the electroneutrality constraint[27], the CPM study with the electrode under constant pressure[28], and the mass-zero constrained approach.[29, 30] Recently, Dufils et al. extended the SR-CPM method using a finite electric field to simulate potential difference[31] or displacement[32] in a single electrode. Beyond the original simple electrostatic description between the electrode charges, the quantum mechanical potentials, including the effect of the penetration and the exchange-correlation, could be incorporated to better simulate the polarisation of the metal or semi-metal electrodes at a given band structure.[33-36]

Despite the rapid advancement of the CPM approach, several critical limitations remain. First, the essence of CPM is the on-the-fly calculation of equilibrium charge distribution $\mathbf{q} \equiv (q_1, q_2, \ldots, q_N)^T$ on the polarizable electrode atoms. Due to the linearity of Maxwell's equation, it could be directly solved as a matrix equation of $\mathbf{Eq} = \mathbf{b}$ where $\mathbf{E}$ is the Ewald interaction matrix and $\mathbf{b}$ is a known vector (See details in **Sec. IIIA.**). For extensive systems, since computing $\mathbf{E}$ and solving the matrix question $\mathbf{Eq} = \mathbf{b}$ are very time-consuming, the equilibrium charge is usually solved using the matrix-*free* iterative approaches instead, e.g., the Car-Parrinello dynamics in Siepmann and Sprik's work[10], and the conjugate gradient method (CG) used by Reed et al.[11] However, all current matrix or matrix free numerical methods used in CPM have unsatisfactory computing efficiency. Second, the value of $\alpha_i$ (the width of the Gaussian charge) could significantly change the polarizability of the conducting surfaces.[24, 34] But no works has ever reported a liable method to determine $\alpha_i$. At present, the choice of $\alpha_i$ is arbitrary and empirical.

To address these issues, in the first part of this work, we develop two new methods to significantly improve the computation efficiency for equilibrium charge $\mathbf{q}$ involved in both the matrix and matrix-free schemes. We reformulate the mathematical framework of the Ewald summation method for computing electrostatic energy with an extension to the Gaussian charge (**Sec. II**). We then proposed an efficient and universal method to calculate $\mathbf{E}$ via using available optimised numerical algorithm libraries (**Sec. IIIB.1**). As for solving $\mathbf{Eq} = \mathbf{b}$, we improve the rate of convergence in the CG method remarkably via the preconditioning techniques (PCG) as $\mathbf{PEq} = \mathbf{Pb}$ (**Sec. IIIB.2**). Note that the obtained knowledge of $\mathbf{E}$ enables us to construct a high-quality sparse preconditioner $\mathbf{P}$ with the optimal convergence rate. Our PCG method can handle the extra-large system with over 8.1 million electrode atoms. In the second part, we developed a method to determine the $\alpha_i$ values based on clear physical principles. The $\alpha_i^0$ for some typical 2D lattices (Table 1) are derived (**Sec. IIIC**).

Our paper is organised as follows. **Section II** provides a reformulated description of the theoretical background for the Ewald summation and the pairwise Ewald potential for both point and Gaussian charges. **Section III** introduces our extended development for SR-CPM regarding the theoretical framework and the numeric implementation, *e.g.*, the electroneutrality constraint, a general computational method for $\mathbf{E}$, the PCG approach to solve the $\mathbf{Eq} = \mathbf{b}$ problem, and the method to determine the Gaussian charge parameter $\alpha_i$. **Section IV** discusses the computational efficiency of our developed methods and the verification of our methods. Finally, we conclude in **Section V**.

## II. Overview of Ewald summation

The long-range nature of the Coulomb interaction leads to the difficulty of electrostatic calculation for systems with massive particles or with a complicated charge distribution[37] For example, the electrostatic binding energy of the rock salt, *i.e.*, NaCl crystal, cannot be solved by a direct sum of the pairwise Coulomb potential between the cation and anion, since the summation result is conditionally convergent.[38, 39] The Ewald summation, proposed in 1921, is the most widely used method to solve the binding energy of the ionic crystal under the 3D periodic boundary conditions (PBC).[40] A plenty of related methods have been proposed to handle and accelerate the electrostatic calculation under diverse conditions, *e.g.*, Lerkner summatifon[41], Mean Many-body Method in 2D (MMM2D)[42], low dimension Ewald summation[43, 44] and Spectral Ewald methods [45] for 1D/2D PBCs, Fast Multipole Method (FMM) for the interaction in vacuum [46] and induced charge computation method (ICC) for the simulation with the dielectric boundary[17, 18].

The Ewald summation and its numerous derivations have become the *de facto* standard in the modern atomistic simulation package.[47] It partitions the energy into the near (real) and far (reciprocal) parts. In practice, the mesh-based Ewald summation method using the Faster Fourier Transform (FFT) technique successfully reduces the time complexity from the original $\mathcal{O}(N^2)$ to $\mathcal{O}(N\log N)$, including the methods of Particle-Particle-Particle-Mesh (P3M)[48], particle mesh Ewald (PME)[49], smooth particle mesh Ewald (SPME)[50], Gaussian split Ewald (GSE) method[51], non-equispaced FFT based fast Ewald summation[52, 53] and Ewald spectral methods[45, 54]

Most Ewald summation methods aim to estimate the interaction energy among point charges, given by a "*Dirac*" or delta distribution. Such a distribution deviates considerably from the realistic electron cloud surrounding the nucleus. The singular self-energy of a point charge also makes it difficult to evaluate the interaction at an extremely short distance. As a result, almost all generic force fields for organic or biological giant molecules use empirical coefficients to screen the intramolecular interactions of bonded atoms.[55-58] Accordingly, the point charge is frequently recommended to be replaced by a more realistically distributed one to avoid singularity.

Due to its computational simplicity and analyticity, the Gaussian function (distribution) is broadly employed in atomistic modelling and quantum chemistry research. A set of Gaussian orbitals, for example, could construct the slater-type atomic orbits easily. [59] Hence, the spherical Gaussian distributed charge (the *Gaussian charge*) has been widely applied in the atomic simulation for various materials, *e.g.*, the water molecule[60], the Metal-Organic Frameworks[61] and the giant biomolecules[62]. To calculate the electrostatic interaction in CPM, the Ewald summation should be extended to cover the interaction of the Gaussian charge. Since the Gaussian charge is deeply involved in the original Ewald summation technique, the extensions to include the Gaussian charge are straightforward.[60, 63]

## A. Prerequisite

We start with the interactions between point and Gaussian charges in both real and reciprocal spaces.[60] The electrostatic potential or energy between two unit point charges $q_i$ and $q_j$ are inversely proportional to the separation distance as $r = |\mathbf{r}|$. Note that we use the Gaussian system of units throughout the paper. As a result, the conventional $1/4\pi\varepsilon_0$ factor is dropped in the following equation,

$$\varphi(\mathbf{r}) = \frac{1}{r} \tag{1}$$

where **r** is the distance vector between two charges. Eq. (1) is also known as the *Green's* function for point charge in a 3D free space. By taking the Fourier transform of $\mathcal{F}\{\varphi\}$, Green's function in reciprocal space yields (See details in the Supplementary Note).

$$\hat{\varphi}(\mathbf{k}) = \mathcal{F}\{\varphi\} = \int \frac{1}{r} e^{i\mathbf{k}\cdot\mathbf{r}} d\mathbf{r} = \frac{4\pi}{k^2} \tag{2}$$

where $k$ is the length of the reciprocal vector as $k = |\mathbf{k}|$, and i is the complex imaginary unit. The interaction in real space is taken back via an inverse Fourier transform of $\mathcal{F}^{-1}\{\varphi\}$,

$$\varphi(\mathbf{r}) = \mathcal{F}^{-1}\{\hat{\varphi}(\mathbf{k})\} = \frac{1}{8\pi^3} \int \frac{4\pi}{k^2} e^{-i\mathbf{r}\cdot\mathbf{k}} d\mathbf{k} = \frac{1}{r} \tag{3}$$

Unlike the point charge, the Gaussian charge is a spatially distributed charge with spherical symmetry. The charge distribution is a Gaussian function,

$$G_i(\mathbf{r}) = \frac{1}{(2\pi\sigma_i^2)^{3/2}} \exp\left(-\frac{|\mathbf{r}-\mathbf{r}_i|^2}{2\sigma_i^2}\right) = \left(\frac{\alpha_i^2}{\pi}\right)^{3/2} \exp(-\alpha_i^2|\mathbf{r}-\mathbf{r}_i|^2) \tag{4}$$

where $\mathbf{r}_i$ is the centre of the Gaussian charge, $\sigma_i$ and $\alpha_i = 1/\sqrt{2}\sigma_i$ are the normal and reciprocal distribution widths, respectively. The electrostatic energy between two Gaussian charges $q_i$ and $q_j$ (the *Green*'s function in vacuum) can be expressed via a Coulomb integral as (See details in the Supplementary Note),

$$\varphi(\mathbf{r};\alpha_{ij}) = \iint \frac{G_i(\mathbf{r}_1)G_j(\mathbf{r}_2)}{|\mathbf{r}_1-\mathbf{r}_2|} d\mathbf{r}_1 d\mathbf{r}_2 = \frac{\mathrm{erf}(\alpha_{ij}|\mathbf{r}|)}{|\mathbf{r}|} \tag{5}$$

where $\mathbf{r} = \mathbf{r}_i - \mathbf{r}_j$ is the distance between the centres of two charges and $\mathrm{erf}(z)$ is the monotonic Gaussian error function where $\mathrm{erf}(0) = 0$ and $\mathrm{erf}(\infty) = 1$. Eq. (5) indicates that the Coulomb interaction of Eq. (1) is screened at a short distance but recovers at large separation. $\alpha_{ij}$ is the mixing interaction coefficient that determines from the Gaussian cross-width law[25] as $\alpha_{ij}^{-2} = \alpha_i^{-2} + \alpha_j^{-2}$.

Eq. (5) can be used to derive some useful formulas. For $\alpha_j \to \infty$, the Gaussian charge $q_j$ reduces to a point charge. In this case, $\alpha_{ij} = \alpha_i$. Replacing $\alpha_{ij}$ with $\alpha_i$ in Eq. (5) yields the potential energy equation between a point and a Gaussian charge $\varphi(\mathbf{r};\alpha_i) = \frac{\mathrm{erf}(\alpha_i r)}{r}$. Substituting $\mathbf{r} = \mathbf{0}$ into Eq. (5) yields the self-energy of a Gaussian charge.

$$\varphi(\mathbf{0};\alpha_{ii}) = \frac{2}{\sqrt{\pi}} \alpha_{ii} = \sqrt{\frac{2}{\pi}} \alpha_i \tag{6}$$

where $\alpha_{ii}$ is the interaction coefficient between two Gaussian charges of equal distribution as $\alpha_{ii} = \frac{\sqrt{2}}{2}\alpha_i$. Note that the self-energy of the Gaussian charge is non-singular as a result of the short-range screening.

The interaction between two Gaussian charges in the reciprocal space can be derived through a Fourier transform (See details in the Supplementary Note).

$$\hat{\varphi}(\mathbf{k}; \alpha_{ij}) = \frac{4\pi}{k^2} \exp\left(-\frac{k^2}{4\alpha_{ij}^2}\right) = \hat{\varphi}(\mathbf{k})\Omega^-(\mathbf{k}; \alpha_{ij}) \tag{7}$$

where $\hat{\varphi}(\mathbf{k})$ is a point charge potential in reciprocal space (Eq. (2)) and $\Omega^- \equiv \exp(-k^2/4\alpha_{ij}^2)$ represents the short-range screening effect in reciprocal space. We defined $\Omega^+ \equiv \exp(+k^2/4\alpha_{ij}^2)$ to yield a reverse relation of $\hat{\varphi}(\mathbf{k}) = \hat{\varphi}(\mathbf{k}; \alpha)\Omega^+(\mathbf{k}; \alpha)$ and $\Omega^+\Omega^- = 1$. Namely, we define two transform functions $\Omega^\pm(\mathbf{k})$ to connect the potential of Gaussian charges with that of point charges in the reciprocal space. It is worth noting that the study involving $N$ different Gaussian charges $\alpha_i$ would introduce $(N+1)N/2$ different mixing coefficients $\alpha_{ij}$, which is complicated in an extended system.

## B. Classic Ewald summation

There are many tutorials[64-66], literatures[67-79] and textbooks[39, 47, 80] that discuss the formula of Ewald summation and the corresponding mathematical and physical meaning. Unfortunately, most derivations are overly complicated, containing several indirect contribution terms or sophisticated integral transformations. As a result, it is difficult for a beginner to grasp the idea behind such excellent work and form a clear physical image. Some literatures[64, 71, 79] even stated the Ewald summation as a certain kind of 'black magic'. In this section, we demystify the magic and provide a transparent description of the Ewald summation, only using the newly developed symbols, $\varphi$ and $\Omega^\pm(\mathbf{k})$, proposed in the previous section. To circumvent the inconvenient expressions, we derive the formula based on a bottom-up approach, starting from the pairwise Ewald potential[65], which could be regarded as the counterpart of the Coulomb potential under the PBCs.

Without loss of generality, we begin with a single point charge, $q_i = 1$, located in a cubic periodic cell $(L = L_x = L_y = L_z)$ at the position of $\mathbf{r}_i$. Then, $q_i$ with all images in the replicated cells form a Simple Cubic (SC) lattice of length $L$. Thus, the electrostatic potential at an arbitrary point $\mathbf{r}$ by $q_i$ is given by the following lattice summation,

$$\psi_i(\mathbf{r}) = {\sum_{\mathbf{n}}}' \frac{q_i}{|\mathbf{r} - \mathbf{r}_i - \mathbf{R}_\mathbf{n}|} \tag{8}$$

where $\mathbf{R_n} = \mathbf{n} \circ \mathbf{L}$ is the cell propagator vector (or simply speaking, the SC lattice points) and $\circ$ is the Hadamard product with $\mathbf{n} = (n_x, n_y, n_z)$, and $n_x, n_y, n_z \in \mathbb{Z}$. The prime in the summation indicates the exclusion of $\mathbf{n} = \mathbf{0}$ at $\mathbf{r} = \mathbf{r}_i$. The long-range nature of Coulomb's interaction causes a divergent summation in Eq. (8). Thus, a uniform neutralising charge background is commonly added to counterbalance such a divergent summation,[65, 73]

$$\psi_i(\mathbf{r}) = \sum_{\mathbf{n}}{}' \frac{q_i}{|\mathbf{r} - \mathbf{r}_i - \mathbf{R_n}|} - \int_{\mathbb{R}^3} \frac{\rho_i}{|\mathbf{r} - \mathbf{r}'|} d\mathbf{r}' \qquad (9)$$

where $\rho_i = q_i/V$ is the density of the background and $V = L^3$ is the volume of the simulation cell. Under the electroneutrality condition ($Q \equiv \sum_i q_i = 0$), the contributions of those backgrounds from each point charge would be cancelled with each other. Eq. (9) itself embodies mathematical and physical significance. It is an analytic continuation of the polytropic potential $\varphi(\mathbf{r}; s) = |\mathbf{r}|^{-2s}$ at $s = 1/2$ in an SC lattice summation.[40, 81] It also describes the energy of the real physical system, such as the Wigner crystal from condensed matter physics[82] or the Coulomb crystal from plasma physics[83], in which the gas of electrons or charged ions/particles condensates in a static, uniformly oppositely-charged background. For simplicity, let us put $q_i$ at the origin point, *i.e.*, $\mathbf{r} = \mathbf{0}$. The resultant potential function is $\psi(\mathbf{r}) = \psi_i(\mathbf{r} - \mathbf{r}_i)$. Since $\psi(\mathbf{r})$ *implicitly* depends on the position of $q_i$, the subscript $i$ is omitted for brevity, and the symbol of $q_i$ is also removed for $q_i = 1$.

To help understand the screening effect of $\rho_i$, we convert the homogeneous background into a compensating Gaussian charge centred at the same position as $q_i$. It yields a two lattice summation formula as

$$\begin{aligned} \psi(\mathbf{r}) &= \sum_{\mathbf{n}}{}' \left[ \frac{1}{|\mathbf{r} - \mathbf{R_n}|} - \frac{\mathrm{erf}(\lambda|\mathbf{r} - \mathbf{R_n}|)}{|\mathbf{r} - \mathbf{R_n}|} \right] \\ &= \sum_{\mathbf{n}}{}' \frac{\mathrm{erfc}(\lambda|\mathbf{r} - \mathbf{R_n}|)}{|\mathbf{r} - \mathbf{R_n}|} \qquad (10) \\ &= \sum_{\mathbf{n}}{}' [\varphi_{\mathbf{n}}(\mathbf{r}) - \varphi_{\mathbf{n}}(\mathbf{r}; \lambda)] \end{aligned}$$

Compared with Eq. (9), the integral of the background in the 2$^{\text{nd}}$ term is replaced by a lattice summation of $\varphi(\mathbf{r}, \lambda)$. The summand $\phi(\mathbf{r}) = \varphi(\mathbf{r}) - \varphi(\mathbf{r}, \lambda)$ represents a screened Coulomb potential, *i.e.* the potential difference between a point charge and its associated compensating Gaussian charge. $\lambda$ is the inverse width of the neutralising Gaussian charge that controls screening strength. $\text{erfc}(z)$ is complementary error function as $\text{erfc}(z) = 1 - \text{erf}(z)$, where $\text{erfc}(0) = 1$ and $\text{erfc}(\infty) = 0$. For $\lambda \to \infty$, the point charge $q_i$ and its periodic images are completely screened by the overlapped point charges. Oppositely, the homogeneous background is achieved at $\lambda \to 0$, like Eq. (9). Note that the introduction of the Gaussian screening function yields an absolute convergent result at arbitrary $\lambda$. The other form of screening function or convergent factor, like Yukawa potential, has ever been introduced in the derivation of the Ewald summation.[42, 73] The Yukawa potential describes the screening charge acting as the ionic electrolyte around the charged particles.

The $\psi(\mathbf{r})$ in Eq. (9) is known as the Ewald potential in MD simulations.[65] It was initiated by Born's *Grundpotential* idea[39] and made explicitly by Emersleben[84] and Hund[85]. The Ewald potential is periodic along with the simulation cell $\psi(\mathbf{r}) = \psi(\mathbf{r} + \mathbf{R_n})$. Hence, the representation in the reciprocal space $\hat{\psi}(\mathbf{k})$ is zero everywhere except at the reciprocal lattice points $\mathbf{K}$. (See details in the Supplementary Note)

$$\hat{\psi}(\mathbf{k}) = \frac{8\pi^3}{V} \hat{\phi}(\mathbf{k}) \sum_{\mathbf{K}} \delta(\mathbf{k} - \mathbf{K})$$
$$= \frac{8\pi^3}{V} [\hat{\varphi}(\mathbf{k}) - \hat{\varphi}(\mathbf{k}; \lambda)] \sum_{\mathbf{K}} \delta(\mathbf{k} - \mathbf{K}) \qquad (11)$$
$$= \frac{8\pi^3}{V} \frac{4\pi}{k^2} \left[1 - \exp\left(-\frac{k^2}{4\lambda^2}\right)\right] \sum_{\mathbf{K}} \delta(\mathbf{k} - \mathbf{K})$$

where $\mathbf{K} = \left(\frac{2\pi n_x}{L}, \frac{2\pi n_y}{L}, \frac{2\pi n_z}{L}\right)$ is the reciprocal lattice point or the reciprocal propagator vector. $\delta(\mathbf{k}) = \delta(k_x)\delta(k_y)\delta(k_z)$ is the 3D Dirac delta function, and $\sum_{\mathbf{K}} \delta(\mathbf{k} - \mathbf{K})$ is the 3D *Dirac Comb*. Note that the factor of $8\pi^3/V$ is the volume of the 1st Brillouin zone. Here, $\hat{\phi}(\mathbf{k}) = \hat{\varphi}^P(\mathbf{k}) - \hat{\varphi}^G(\mathbf{k}; \lambda)$ describes a potential difference between a point charge and its associated compensating Gaussian charge in the reciprocal space.

The Ewald summation partitions the electrostatic potential $\hat{\phi}$ into two parts, a short-range part $\hat{\phi}^S \equiv (1-d)\hat{\phi}$ and a long-range part $\hat{\phi}^L \equiv d\hat{\phi}$, respectively. The choice of the partition function $d$ is generally arbitrary (See Page 23 in Ref. 39). Many different kinds of partition functions have been proposed either in the real or in the reciprocal spaces.[86-89] Herein, we present the original Gaussian partition function in the reciprocal space using the newly proposed symbol of $\Omega^-(\mathbf{k};\eta)$,

$$\hat{\phi}^S(\mathbf{k}) = \hat{\varphi}(\mathbf{k})[1 - \Omega^-(\mathbf{k},\eta)]$$
$$= \hat{\varphi}(\mathbf{k}) - \hat{\varphi}(\mathbf{k};\eta) = \frac{4\pi}{k^2}\left[1 - \exp\left(-\frac{k^2}{4\eta^2}\right)\right]$$

$$\hat{\phi}^L(\mathbf{k}) = \hat{\varphi}(\mathbf{k})\Omega^-(\mathbf{k},\eta) - \hat{\varphi}(\mathbf{k};\lambda) = \begin{cases} \frac{4\pi}{k^2}\exp\left(-\frac{k^2}{4\eta^2}\right), & \mathbf{k} \neq \mathbf{0}, \lambda \to 0 \\ -\frac{\pi}{\eta^2}, & \mathbf{k} = \mathbf{0}, \lambda \to 0 \end{cases} \quad (12)$$

Note that the short-range part of $\hat{\phi}^S(\mathbf{k})$ is like the previous $\hat{\phi}(\mathbf{k})$ except for the different factors $\lambda$ vs $\eta$. The latter $\eta$ is the screening factor in the Ewald summation, which is usually comparable to the dimension of the reciprocal cell, $L^{-1}$, in contrast to $\lambda \to 0$ for the neutralising background. Correspondingly, the long-range part of $\hat{\phi}^L(\mathbf{k})$ can be viewed from the potential of the Gaussian charge in the reciprocal space $\hat{\varphi}(\mathbf{k};\eta) = \hat{\varphi}(\mathbf{k})\Omega^-(\mathbf{k};\eta)$ minus by the neutralising term $\hat{\varphi}(\mathbf{k};\lambda)$. Considering $\lambda \to 0$, $\hat{\varphi}(\mathbf{k};\lambda)$ is vanishing everywhere except at $\mathbf{k} = (0,0,0)$, where the value of $\hat{\phi}^L$ at that point is obtained from $\lim_{k\to 0}\frac{4\pi}{k^2}(e^{-k^2/4\eta^2} - 1) = -\frac{\pi}{\eta^2}$. With the partition of $\hat{\phi}(\mathbf{k})$, we derive the tractable Ewald potential, using an inverse Fourier transform,

$$\psi(\mathbf{r}) = \sum_\mathbf{n} \frac{\mathrm{erfc}(\eta|\mathbf{r}-\mathbf{R_n}|)}{|\mathbf{r}-\mathbf{R_n}|} + \frac{4\pi}{V}\sum_{\mathbf{K}\neq 0}\frac{e^{-i\mathbf{K}\cdot\mathbf{r}}}{K^2}\exp\left(-\frac{K^2}{4\eta^2}\right) - \frac{\pi}{V\eta^2} - \lim_{\mathbf{r}\to 0}\frac{1}{|\mathbf{r}|} \quad (13)$$

The 1$^{st}$ term is the Fourier transform of the short-range components $\phi^S$, leading to a real-space summation term truncated by $\mathrm{erfc}(z)$. Likewise, the 2$^{nd}$ term is derived from the integral with the Dirac comb in Eq. (11), which could be handled efficiently by the exponential decay in the reciprocal space. The 3$^{rd}$ term is just $\hat{\phi}^L(\mathbf{0})$ multiplied by $V^{-1}$. The last term is added to exclude the singular self-interaction at $\mathbf{n} = \mathbf{0}$ and $\mathbf{r} = \mathbf{0}$. The screening effect by $\eta$, yielding a convergent summation in real space, is corrected by the 2$^{nd}$ term in reciprocal space, and the 3$^{rd}$ term accounts for the non-neutral contribution. Based on the Ewald potential in Eq. (13), the total electrostatic energy of $N$ different point charges in a periodic cell is assembled as

$$U = \frac{1}{2} \int_V \sum_i^N q_i \, \delta(\mathbf{r} - \mathbf{r}_i) \sum_j^N q_j \, \psi_j(\mathbf{r}) d\mathbf{r}$$

$$= \frac{1}{2} \sum_{i,j} q_i q_j \left\{ \sum_{\mathbf{n}}' \frac{\text{erfc}(\eta|\mathbf{r}_{ij} - \mathbf{R}_\mathbf{n}|)}{|\mathbf{r}_{ij} - \mathbf{R}_\mathbf{n}|} + \frac{4\pi}{V} \sum_{\mathbf{K} \neq 0} \frac{e^{-i\mathbf{K} \cdot \mathbf{r}_{ij}}}{K^2} \exp\left(-\frac{K^2}{4\eta^2}\right) \right\} \quad (14)$$

$$- \frac{\pi}{2V\eta^2} Q^2 - \frac{\eta}{\sqrt{\pi}} \sum_i q_i^2$$

where $\mathbf{r}_{ij}$ is the distance between two charges as $\mathbf{r}_{ij} = \mathbf{r}_i - \mathbf{r}_j$ and the factor of $\eta/\sqrt{\pi}$ in the last term is taken from $\lim_{r \to 0} \text{erf}(\eta r)/r = 2\eta/\sqrt{\pi}$, namely the self-energy of a Gaussian charge, $\varphi(\mathbf{0}; \eta) = 2\eta/\sqrt{\pi}$. Note that our derivation is the same under the tinfoil boundary condition,[77, 90] because the associated compensating Gaussian charges remove the contribution of the dipole moment in the reciprocal space ($\mathbf{k} = \mathbf{0}$).

According to the form of Eq. (14) and the linearity of Maxwell's equations, the total electrostatic energy could be expressed using a linear electrostatic functional of $\mathbf{q}$ as $U\{\mathbf{q}\} = \frac{1}{2} \mathbf{q}^T \mathbf{E} \mathbf{q}$ where $\mathbf{q} = [q_1, q_2, \ldots, q_N]^T$ is the (charge) configuration vector, including all the charges on each atom, and $\mathbf{E}$ is the symmetric Ewald interaction matrix. Thus, the essential work of Ewald summation is to directly or indirectly determine each entry $e_{ij}$ of $\mathbf{E}$ as

$$e_{ij} = \psi_j(\mathbf{r}_i) = \psi^0(\mathbf{r}_{ij}) - \frac{\pi}{V\eta^2} + \begin{cases} \frac{\text{erfc}(\eta|\mathbf{r}_{ij}|)}{|\mathbf{r}_{ij}|}, & \mathbf{r}_i \neq \mathbf{r}_j \\ -\frac{2\eta}{\sqrt{\pi}}, & \mathbf{r}_i = \mathbf{r}_j \end{cases} \quad (15)$$

where $\psi^0(\mathbf{r})$ is defined as:

$$\psi^0(\mathbf{r}) = \sum_{\mathbf{n} \neq \mathbf{0}} \varphi_\mathbf{n}(\mathbf{r}) - \varphi_\mathbf{n}(\mathbf{r}; \eta) + \frac{1}{V} \sum_{\mathbf{K} \neq 0} \hat{\varphi}(\mathbf{K}; \eta) e^{-i\mathbf{K} \cdot \mathbf{r}} \quad (16)$$

Figure 1 shows the constitution of $\mathbf{E}$ for three-point charges in a cubic periodic cell with different diagonal and off-diagonal elements. A detailed discussion of the structure $\mathbf{E}$ in CPM is given in the Supplementary Note.

## C. Ewald's summation for Gaussian charges

Once the classic Ewald summation is derived, the extension to the Gaussian charge is straightforward. A series of theoretical works were taken to formulate the Gaussian Ewald summation equation under the 2D/3D PBCs.[10, 11, 24, 27, 34, 60, 63, 91] Herein, we reformulate the energy functional in the literature from the pairwise Gaussian Ewald potential, only using the new proposed symbols of $\varphi$ and $\Omega^{\pm}(\mathbf{k})$. In a cubic periodic cell, we can investigate the interaction energy $\psi(\mathbf{r})$ on a probe Gaussian charge $q_j$ from another unit Gaussian charge $q_i$, giving

$$\psi(\mathbf{r}_{ij}; \alpha_{ij}) = \sum_{\mathbf{n}}' \left[ \frac{\text{erf}(\alpha_{ij}|\mathbf{r} - \mathbf{R}_\mathbf{n}|)}{|\mathbf{r}_{ij} - \mathbf{R}_\mathbf{n}|} - \frac{\text{erf}(\lambda|\mathbf{r}_{ij} - \mathbf{R}_\mathbf{n}|)}{|\mathbf{r}_{ij} - \mathbf{R}_\mathbf{n}|} \right] \qquad (17)$$
$$= \sum_{\mathbf{n}}' [\varphi_\mathbf{n}(\mathbf{r}_{ij}; \alpha_{ij}) - \varphi_\mathbf{n}(\mathbf{r}_{ij}; \lambda)]$$

where $\mathbf{r}_{ij} = \mathbf{r}_i - \mathbf{r}_j$ is the distance vector between two Gaussian charges at $\mathbf{r}_i$ and $\mathbf{r}_j$. $\alpha_{ij}$ is the mixing interaction coefficient. The 2$^{\text{nd}}$ term is the neutralising background for $q_i$ with $\lambda \rightarrow 0$, like Eq. (10). Note that the symbols of $q_i$ and $q_j$ are dropped $q_i = q_j = 1$. For simplicity, the subscript of the interaction index $ij$ is omitted in the following equations. Due to the same lattice periodicity, the Gaussian Ewald potential in reciprocal space is similar to that of Eq. (11),

$$\hat{\psi}(\mathbf{k}; \alpha) = \frac{8\pi^3}{V} \hat{\varphi}(\mathbf{k}; \alpha) \sum_{\mathbf{K}} \delta(\mathbf{k} - \mathbf{K}) \qquad (18)$$
$$= \frac{8\pi^3}{V} [\hat{\varphi}(\mathbf{k}; \alpha) - \hat{\varphi}(\mathbf{k}; \lambda)] \sum_{\mathbf{K}} \delta(\mathbf{k} - \mathbf{K})$$

By comparing with Eq. (11), we found $\hat{\varphi}(\mathbf{k})$ is replaced by its Gaussian counterpart. As follows, a different partition function of $\Omega^+(\mathbf{k}, \alpha)\Omega^-(\mathbf{k}, \eta)$ rather than $\Omega^-(\mathbf{k}, \eta)$ is applied,

$$\begin{aligned}
\hat{\varphi}^S(\mathbf{k}; \alpha) &= \hat{\varphi}(\mathbf{k}; \alpha)[1 - \Omega^+(\mathbf{k}, \alpha)\Omega^-(\mathbf{k}, \eta)] \\
&= \hat{\varphi}(\mathbf{k}; \alpha) - \hat{\varphi}(\mathbf{k}; \eta) = \frac{4\pi}{k^2}\left[\exp\left(-\frac{k^2}{4\alpha^2}\right) - \exp\left(-\frac{k^2}{4\eta^2}\right)\right] \\
\hat{\varphi}^L(\mathbf{k}; \alpha) &= \hat{\varphi}(\mathbf{k}; \alpha)\Omega^+(\mathbf{k}, \alpha)\Omega^-(\mathbf{k}, \eta) - \hat{\varphi}(\mathbf{k}; \lambda) \\
&= \hat{\varphi}(\mathbf{k}; \eta) - \hat{\varphi}(\mathbf{k}; \lambda) = \begin{cases} \frac{4\pi}{k^2}\exp\left(-\frac{k^2}{4\eta^2}\right), & \mathbf{k} \neq \mathbf{0}, \lambda \rightarrow 0 \\ -\frac{\pi}{\eta^2}, & \mathbf{k} = \mathbf{0}, \lambda \rightarrow 0 \end{cases}
\end{aligned} \qquad (19)$$

It is worth noting that the long-range part of $\hat{\phi}^L(\mathbf{k}; \alpha)$ becomes the same as that in the classic Ewald summation Eq. (12). Note that $\hat{\varphi}^G(\mathbf{k}; \lambda)$ is zero everywhere except $\mathbf{k} \to \mathbf{0}$ at $\lambda \to 0$. Kiss et al. stated that the advantage of such a partition function is that it simplifies the calculation in the reciprocal space.[60] For example, there are three different $\alpha_{ij}$ in the mixture system (*e.g.*, point-to-point, point-to-Gaussian, and Gaussian-to-Gaussian), resulting in the different $\phi^L(\mathbf{k}; \alpha)$. The partition function unifies the calculation in the reciprocal space as an $\alpha$-independent one. It is thus overwhelmingly advantageous to reuse the existing highly optimised electrostatic library designed for a point charge. The tractable Gaussian Ewald potential is derived using an inverse Fourier transform,

$$\psi(\mathbf{r}) = \sum_{\mathbf{n}} \frac{\text{erfc}(\eta|\mathbf{r}_{ij} - \mathbf{R_n}|)}{|\mathbf{r}_{ij} - \mathbf{R_n}|} - \frac{\text{erfc}(\alpha_{ij}|\mathbf{r}_{ij} - \mathbf{R_n}|)}{|\mathbf{r}_{ij} - \mathbf{R_n}|} + \frac{4\pi}{V} \sum_{K \neq 0} \frac{e^{-i\mathbf{K} \cdot \mathbf{r}}}{K^2} \exp\left(-\frac{K^2}{4\eta^2}\right) - \frac{\pi}{V\eta^2} \quad (20)$$

The total electrostatic energy by $N$ different Gaussian charges in a periodic cell is assembled accordingly as,

$$U = \frac{1}{2} \sum_{i,j} q_i q_j \left\{ \sum_{\mathbf{n}} \frac{\text{erfc}(\eta|\mathbf{r}_{ij} - \mathbf{R_n}|)}{|\mathbf{r}_{ij} - \mathbf{R_n}|} - \frac{\text{erfc}(\alpha_{ij}|\mathbf{r}_{ij} - \mathbf{R_n}|)}{|\mathbf{r}_{ij} - \mathbf{R_n}|} + \frac{4\pi}{V} \sum_{K \neq 0} \frac{e^{-i\mathbf{K} \cdot \mathbf{r}_{ij}}}{K^2} \exp\left(-\frac{K^2}{4\eta^2}\right) \right\} \\ - \frac{\pi}{2V\eta^2} Q^2 - \frac{1}{\sqrt{\pi}} \sum_i q_i^2 (\eta - \alpha_{ii}) \quad (21)$$

where $\mathbf{E}$ is the Gaussian Ewald interaction matrix. Each entry $e_{ij}$ depends on the separation vector $\mathbf{r}_{ij}$ and their mixing interaction coefficient $\alpha_{ij}$.

$$e_{ij}(\alpha_{ij}) = \psi^0(\mathbf{r}_{ij}; \alpha_{ij}) - \frac{\pi}{V\eta^2} + \begin{cases} \dfrac{\text{erfc}(\eta|\mathbf{r}_{ij}|)}{|\mathbf{r}_{ij}|} - \dfrac{\text{erfc}(\alpha_{ij}|\mathbf{r}_{ij}|)}{|\mathbf{r}_{ij}|}, & \mathbf{r}_i \neq \mathbf{r}_j \\ -\dfrac{2\eta}{\sqrt{\pi}} + \dfrac{2\alpha_{ii}}{\sqrt{\pi}}, & \mathbf{r}_i = \mathbf{r}_j \end{cases} \quad (22)$$

where $\psi^0(\mathbf{r}; \alpha)$ is defined as,

$$\psi^0(\mathbf{r}; \alpha) = \sum_{\mathbf{n} \neq 0} \varphi_{i,\mathbf{n}}(\mathbf{r}; \alpha) - \varphi_{i,\mathbf{n}}(\mathbf{r}; \eta) + \frac{4\pi}{V} \sum_{K \neq 0} \hat{\varphi}(\mathbf{K}; \eta) e^{-i\mathbf{K} \cdot \mathbf{r}} \quad (23)$$

Note that the collapse of the Gaussian charge into a point counterpart, i.e., $\alpha \to \infty$ yields the same formula as the classic Ewald summation, except for the diagonal entries where the singular self-energy term of $\sqrt{2/\pi}\alpha_i$ should be removed.

The reformulation of Ewald summation provides a consistent computational framework for both point and Gaussian charges, especially for the more complicated reciprocal space.[11, 24, 60]. Hence, the existing highly optimised electrostatic library, e.g., ScaFaCoS[92], or the internal electrostatic module in the atomistic simulation packages, could be reused with minor modification on the short-range part as well as the hitherto error estimation methods[67, 93]. Moreover, it is also applicable to evaluate the interactions between the other forms of charge distributions, either spherical symmetric or non-symmetric, given that a Gaussian charge could roughly screen the interaction in real space at $r_{\text{cut}}$. Recently, Ahrens-Iwers and Meißner have demonstrated that the CPM problem can be solved with the assistance of P3M[91], which is also the primary method used in our LAMMPS implementation.

## III Further development of CPM

### A. Theoretical framework

The SR-CPM is proposed to model the electrode with a constant potential constraint in electrochemical applications, including supercapacitor, water desalination, and ion transistor. The simulation cell usually consists of mobile electrolyte molecules/ions and immobile electrode atoms. A polarizable electrolyte/electrode interface can be simulated by dynamically changing the charges on electrode atoms. According to the definition in the previous section, the total electrostatic energy (functional) in a CPM study is defined as,

$$U\{\mathbf{q}^{\text{ele}}, \mathbf{q}^{\text{sol}}\} = \frac{1}{2}\sum_{I,J} q_I q_J e_{IJ} + \sum_{i,I} q_i q_I e_{iI}(\alpha_i) + \frac{1}{2}\sum_{i,j} q_i q_j e_{ij}(\alpha_{ij}) - \sum_i \chi_i q_i \quad (24)$$

The capital subscripts represent the atomic charges on the electrolyte ions and molecules $q_I$, and the lower alphabet subscripts are for the electrode atomic charge $q_i$. $\mathbf{q}^{\text{ele/sol}}$ represents the charge configuration vector for the electrode atoms or electrolyte ions/molecules. The first three terms in Eq. (24) describe the electrolyte-electrolyte, electrolyte-electrode, and electrode-electrode interactions, respectively. In the CPM simulation, the electrode charge of $q_i$ is treated as the Gaussian charges with an inverse width of $\alpha_i$, in contrast to the dimensionless electrolyte charges $q_I$. Thus, only the coefficients in the 2$^{nd}$ and 3$^{rd}$ terms are $\alpha$-dependent. The last term is added to describe the given potential constraint $\chi_i$ at $i$th electrode atom. The electrode atomic charges ($\mathbf{q}^{\text{ele}}$) would fluctuate and redistribute to minimise the energy functional Eq. (24). The equilibrium $\mathbf{q}$ can be solved by taking $\partial U / \partial q_i = 0$, namely the chemical potential equalization[34], yielding

$$\sum_j q_j e_{ij}(\alpha_{ij}) = \chi_i - \sum_I q_I e_{iI}(\alpha_i) = \chi_i - b_i^{\text{sol}} = b_i \tag{25}$$

where $b_i^{\text{sol}} \equiv \sum_I q_I e_{iI}(\alpha_i)$ is the electrostatic potential on the $i$th electrode atom from all charges of the electrolyte molecules and ions. Accordingly, we solve it as a linear equation system, $\mathbf{E}\mathbf{q}^{\text{ele}} = \mathbf{b}$, where $\mathbf{E}$ is the (sub-) Ewald interaction matrix between the electrode atoms, and $\mathbf{b}$ is the extrinsic potential vector defined as $b_i \equiv \chi_i - \sum_I q_I e_{iI}$. Namely, $\mathbf{b}$ depends on the electrolyte molecule's spatial configuration and the potential constraint $\chi_i$. In practice, the change of the electrode charge $\Delta q_i$ between the two adjacent MD steps is solved as,

$$\sum_j \Delta q_j e_{ij}(\alpha_{ij}) = \chi_i - \sum_I q_I e_{iI}(\alpha_i) - \sum_j q_j^0 e_{ij}(\alpha_{ij}) \tag{26}$$

where $q_i^0$ is the equilibrium charge in the previous MD step and $\Delta q_i$ is the change of $q_i$ in the following MD step as $q_i = \Delta q_i + q_i^0$. Note that in either Eq. (25) or Eq. (26), the electrolyte-electrode coefficients $e_{iI}$ change with the movement of the electrolyte molecules and ions. We can calculate the influence of the electrolyte, $\mathbf{b}^{\text{sol}}$, using the matrix-*free* Ewald summation method as Eq. (21). On the other hand, the invariant electrode-electrode coefficients $e_{ij}$ pave the way to accelerate the solving of $\mathbf{Eq} = \mathbf{b}$, either from the precalculated $\mathbf{E}^{-1}$ or from the preconditioning technique in iterative approaches (details in **Sec. IIIB**).

During solving Eq. (25) or Eq. (26), we observed the temporal breakdown of the global electroneutrality and the unequal charge quantities on the cathode and anode, which is also observed in the previous study.[91] Such phenomena could be explained from the thermodynamic perspective. From Eq. (24), the Gibbs free energy of the simulation cell under the potential constraint is derived,

$$F = -TS + PV + \sum_{s} \mu_s N_s - \sum_{i} q_i \chi_i \qquad (27)$$

where $\mu_s$ and $N_s$ are the chemical potential and particle number for $s$th species, respectively. Note that the electrode potential constraint $\chi_i$ and the partial charge on electrode $q_i$ are conjugated. For a limited simulation system, the fixed potential constraint inevitably results in a fluctuation of $q_i$. Thus, the global electroneutrality could be temporarily broken down and introduces some artifacts into the Ewald summation, *e.g.*, the homogeneous neutralising background and the removal of charges from the simulation cell. Hence, a Lagrange undetermined multiplier $\mu_c$, *i.e.*, electroneutrality constraint, is added to eliminate the net charge fluctuation[27, 34] as,

$$U\{\mathbf{q}^{\text{ele}}, \mathbf{q}^{\text{sol}}\} = \frac{1}{2}\sum_{I,J} q_I q_J e_{IJ} + \sum_{i,I} q_i q_I e_{iI}(\alpha_i) + \frac{1}{2}\sum_{i,j} q_i q_j e_{ij}(\alpha_{ij}) - \sum_{i} \chi_i q_i + \mu_c Q^{\text{sum}} \qquad (28)$$

where $Q^{\text{sum}} = Q^{\text{ele}} + Q^{\text{sol}}$ is the net charge in the simulation cell, and $Q^{\text{ele/sol}}$ are the total charges on the electrode or the electrolyte molecules as $Q^{\text{ele/sol}} \equiv \sum q_{i/I}$. Taking the derivative of Eq. (28) to $q_i$ yields a shifted linear equation system as

$$\mathbf{Eq} = \mathbf{b} - \mu_c \mathbf{1} \qquad (29)$$

Or explicitly,

$$\sum_{j} q_j e_{ij}(\alpha_{ij}) = \chi_i - \sum_{I} q_I e_{iI} - \mu_c = b_i - \mu_c \qquad (30)$$

where $\mu_c$ describes the amount of the electrochemical potential change and $\mathbf{1} \equiv (1,1,\dots,1)^T$ is an $N$-unit vector. Compared with Eq. (25), the constraint of $\chi_i$ on each electrode atom is shifted by $\mu_c$, but the potential difference (voltage between electrodes) remains invariant. It is worth noting that the electroneutrality constraint changes the chemical potentials of all the charges (both $q_i$ and $q_I$). The simulation is then configurationally sampled under the constant potential difference between the electrodes.[27]

Recently, Scalfi et al. enforced electroneutrality by using a matrix $\mathbf{S}$ where the corresponding $\mathbf{q}^{\text{ele}}$ is obtained simply via $\mathbf{q}^{\text{ele}} = \mathbf{Sb}$. To understand the role of $\mathbf{S}$, we first partition the equilibrium $\mathbf{q}^{\text{ele}}$ into two parts of $\mathbf{q}^b$ and $\mathbf{q}^{\mu_c}$ as

$$\mathbf{q}^{\text{ele}} = \mathbf{q}^b + \mathbf{q}^{\mu_c}$$
$$= \mathbf{E}^{-1}\mathbf{b} - \mu_c \mathbf{E}^{-1}\mathbf{1} \tag{31}$$

where $\mathbf{q}^b$ and $\mathbf{q}^{\mu_c}$ are the charge configurations calculated from $\mathbf{b}$ and $\mathbf{\mu_c}$, respectively. To satisfy the electroneutrality with the conventional electrolyte solution $Q^{\text{sol}} = 0$, the total charge on the electrodes, $Q^{\text{ele}}$, must be zero, i.e., $\mathbf{1}^T \mathbf{q}^{\text{ele}} = 0$, yielding,

$$\mu_c = \frac{\mathbf{1}^T \mathbf{E}^{-1}}{\mathbf{1}^T \mathbf{E}^{-1} \mathbf{1}} \mathbf{b} = \mathbf{d}^T \mathbf{b}$$
$$= \frac{Q^b}{C^{\text{ele}}} \tag{32}$$

where $Q^b \equiv \mathbf{1}^T \mathbf{E}^{-1} \mathbf{b} = \sum_i q_i^b$ is the total charge quantity on all electrodes determined by $\mathbf{b}$ and $C^{\text{ele}} \equiv \mathbf{1}^T \mathbf{E}^{-1} \mathbf{1} = \sum_{i,j} e_{ij}^{-\hat{1}}$ is the total capacitance of the system of electrodes held at the same potential[94], and $e_{ij}^{-\hat{1}}$ is the element in the inversed matrix $\mathbf{E}^{-1}$. As a result, the ratio of charge $Q^b$ over $C^{\text{ele}}$ yields the potential (shift) $\mu_c$ for the electroneutrality constraint. Note that the electric field lines emerging from these electrodes at the same potential would dissolve in the neutralising background charge under the PBCs. $\mathbf{d}$, defined as $d_i = \frac{1}{C^{\text{ele}}} \sum_j e_{ij}^{-\hat{1}}$, is the charge partitioning vector or neutralising vector on the electrode, which equals $\frac{1}{Q^{\mu_c}} \mathbf{q}^{\mu_c}$. It indicates that the relative change of charges by the electroneutrality constraint (held at the same potential) is invariant. Substituting Eq. (32) back into Eq. (31), yielding,

$$\mathbf{q}^{\text{ele}} = \mathbf{E}^{-1}\mathbf{b} - \mathbf{E}^{-1}\mathbf{1}\mathbf{d}^T \mathbf{b}$$
$$= \mathbf{E}^{-1}(\mathbf{I} - \mathbf{1}\mathbf{d}^T)\mathbf{b} \tag{33}$$
$$= \mathbf{S}\mathbf{b}$$

where $\mathbf{I}$ is the $N \times N$ identity matrix. $\mathbf{S}$ is a singular symmetric matrix defined as $s_{ij} = e_{ij}^{-\hat{1}} - d_j \sum_k e_{ik}^{-\hat{1}}$. $\{a\mathbf{1} | a \in \mathbb{R}\}$ is the kernel of $\mathbf{S}$ with $N - 1$ rank as $a\mathbf{S}\mathbf{1} = \mathbf{0}$. Thus, the image of $\mathbf{S}$, i.e., $\mathbf{q}^{\text{ele}}$, always ensures the constraint of $Q^{\text{ele}} = 0$. Indeed, $\mathbf{S}$ describes the polarizability of the cathode and anode, i.e., the capability to redistribute the charges between the electrodes under the potential gradient (electric field), as the charge response kernel (CRK) matrix in the polarizable charge method.[95]

However, the use of $\mathbf{S}$ requires $\mathbf{E}^{-1}$, which prohibits a large-scale simulation study and the iterative techniques used in CPM. Moreover, it is not applied to studying the monopolar electrolyte ($Q^{sol} \neq 0$) in the half-cell configuration (the single electrode).[96] In the monopolar case, $\mu_c$ should satisfy the constraint of $\mathbf{1}^T \mathbf{q}^{ele} + Q^{sol} = 0$, yielding

$$\mu_c = \mathbf{d}^T \mathbf{b} + \frac{Q^{sol}}{\mathbf{1}^T \mathbf{E}^{-1} \mathbf{1}} \tag{34}$$

By substituting Eq. (34) back into Eq. (31) and using the relation of $\mathbf{E}^{-1} \mathbf{1} \mathbf{d}^T = \mathbf{d} \mathbf{1}^T \mathbf{E}^{-1}$, we derive a *two-step* method to derive $\mathbf{q}^{ele}$ in a general simulation cell, as

$$\begin{aligned}\mathbf{q}^{ele} &= \mathbf{E}^{-1} \mathbf{b} - \mathbf{d} \mathbf{1}^T \mathbf{E}^{-1} \mathbf{b} - \mathbf{d} Q^{sol} \\ &= \mathbf{q}^b - \mathbf{d}(\mathbf{1}^T \mathbf{q}^b + Q^{sol}) \\ &= \mathbf{q}^b - (Q^b + Q^{sol})\mathbf{d}\end{aligned} \tag{35}$$

In the first step, $\mathbf{q}^b$ and the total $Q^b$ are computed using $\mathbf{q}^b = \mathbf{E}^{-1} \mathbf{b}$. Then, the system is neutralized in the 2nd post-treatment step by $(Q^b + Q^{sol})\mathbf{d}$. Such treatment is the same as the Lagrangian method applied for Charge equilibrium (Qeq) model[97] where the partial atomic charge $q_i$ has to be solved on-the-fly onto the molecules with the varying structure rather than the fixed electrode atoms in CPM study. Our two-step method is feasible for both the matrix and the iterative techniques used in CPM (See details in the Supplementary Note) and is applicable for the non-neutral electrolyte studies ($Q^{sol} \neq 0$). In addition, the post-treatment step can be employed to handle some other, more complex electrostatic constraints on the fluctuating conducting surface, using the capacitance coefficients $C_{ab}$[98], such as the electrode charge constrain in the Galvanostatic charge-discharging (GCG) process [32, 99] (See details in the Supplementary Note) and the $\mu$NVT ensemble simulation[100]. We will present these study results in other publications.

## B. Numerical techniques

### 1. Matrix inversion (MI)

In CPM, the matrix inversion (MI) technique can update the equilibrium electrode atomic charges $\mathbf{q}^{ele}$ at each MD step, using $\mathbf{q}^{ele} = \mathbf{E}^{-1} \mathbf{b}$.[34, 91]. However, it is time-consuming to calculate $\mathbf{E}$ or $\mathbf{E}^{-1}$ via the currently available method. This section develops a method with a significant improvement in computational efficiency.

Wang et al. calculated **E** element-by-element with an internal Gaussian Ewald summation[4] that results in a biquadratic scaling of $\mathcal{O}(N^4)$ (dashed red curve in Fig. 7a). Recently, Ahrens-Iwers and Meißner remarkably improved efficiency by introducing an intermediate matrix **X** in the mesh-based P3M method.[91] Their implementation employs the established Green's function on the regular grid with $\mathcal{O}(N^2)$ scaling. However, the trade-off is the prohibitive memory cost of **X** and some in-depth and specific modifications of the Mesh-based Ewald code (P3M in that literature). Herein, we propose an alternative approach to calculating **E** using $N$ times of Ewald summation that circumvents the storage of **X**, which is suitable for an extensive system. More importantly, our approach is compatible with all existing highly optimised electrostatic libraries without specific code modification.

Figure 1 is a pedagogical example of calculating **E** for a periodic cubic cell ($L = 1$) with three point-charges of $q_1$, $q_2$ and $q_3$ in a cubic periodic cell. In this case, $\tilde{\mathbf{q}} \equiv (q_1, q_2, q_3)^T$ is the charge vector, and $\tilde{\mathbf{u}} \equiv (u_1, u_2, u_3)^T$ is the potential vector for these three charges. We select three different charge configurations as $\tilde{\mathbf{q}}^1 = (3, -1, -1)^T$, $\tilde{\mathbf{q}}^2 = (-1, 3, -1)^T$ and $\tilde{\mathbf{q}}^3 = (-1, -1, 3)^T$, respectively. Then, we calculate the corresponding potential profiles of $\tilde{\mathbf{u}}^1$, $\tilde{\mathbf{u}}^2$ and $\tilde{\mathbf{u}}^3$ by Eq. (21), using the existing matrix-*free* Ewald summation libraries/codes. The charge and potential matrices are assembled as $\tilde{\mathbf{Q}} \equiv (\tilde{\mathbf{q}}^1, \tilde{\mathbf{q}}^2, \tilde{\mathbf{q}}^3)$ or $\tilde{\mathbf{U}} \equiv (\tilde{\mathbf{u}}^1, \tilde{\mathbf{u}}^2, \tilde{\mathbf{u}}^3)$. Thus, **E** can be calculated from $\mathbf{E} = \tilde{\mathbf{U}}\tilde{\mathbf{Q}}^{-1}$. Although Fig. 1 only illustrates the interactions between point charges, our method can handle arbitrary types of charges. Note that the obtained diagonal element in **E** is the well-established Madelung constant for a (Wigner) SC lattice, $\xi = -2.83729479$[101], which plays a significant role involving the periodic pattern. (See details in the Supplementary Note).

The following steps are the general simulation procedures for an electrode composed of $N$ atoms,

1. An $N \times N$ symmetric invertible charge matrix $\tilde{\mathbf{Q}}$ is selected as $\tilde{\mathbf{Q}} = (\tilde{\mathbf{q}}^1, \tilde{\mathbf{q}}^2, \dots, \tilde{\mathbf{q}}^N)$. Namely, each column $\tilde{\mathbf{q}}^j$ is a specific charge configuration for the $j$th Ewald summation.

2. At $j$th step, the $i$th electrode charge is assigned to be $\tilde{q}_{ij}$, namely the $i$th entry of the $j$th column of $\tilde{\mathbf{q}}^j$.

3. The corresponding potentials on the electrode atoms by $\tilde{\mathbf{q}}^j$ are obtained via an Ewald summation of Eq. (21)

4. To repeat steps (2)-(3) by $N$-times, $N$ different potential vectors $\widetilde{\mathbf{u}}^{\{1,\ldots,N\}}$ are obtained from $N$ individual Ewald summations. Thus, the symmetric matrix $\widetilde{\mathbf{U}}$ is assembled as $\widetilde{\mathbf{U}} \equiv (\widetilde{\mathbf{u}}^1, \widetilde{\mathbf{u}}^2, \ldots, \widetilde{\mathbf{u}}^N)$ with the relation of $\mathbf{E} = \widetilde{\mathbf{U}}\widetilde{\mathbf{Q}}^{-1}$ or $\mathbf{E}^{-1} = \widetilde{\mathbf{Q}}\widetilde{\mathbf{U}}^{-1}$.

The input $\widetilde{\mathbf{Q}}$ is specifically designed with the same diagonal or off-diagonal entries. As our pedagogical example, $\widetilde{\mathbf{Q}}$, defined as $q_{ii} = 3$ and $q_{ij} = -1$, could easily be converted into $\widetilde{q}_{ii}^{-1} = 0.5$ and $\widetilde{q}^{-1} = 0.25$, respectively (See the Supplementary Note). Thus, in our implementation, $\mathbf{E}$ is computed using $\mathbf{E} = \widetilde{\mathbf{U}}\widetilde{\mathbf{Q}}^{-1}$. Then $\mathbf{E}^{-1}$ can be obtained by the inverse of the matrix. Once the FFT techniques with the quasilinear scaling $\mathcal{O}(N\log N)$ are employed in the particle mesh, the overall $N$-times Ewald summations yield a quasi-quadratic time cost in calculating $\mathbf{E}$. Our method is thus comparable with the intermediate matrix method. An example will be demonstrated later. Nevertheless, the matrix inversion step costs at least $N^3$ operations, which could become a bottleneck in an extensive system. We can further accelerate this step by using the distributed numeric library, *i.e.*, ScaLAPACK[102] (Fig. S6).

Our new method is a general yet straightforward approach to determining $\mathbf{E}$ that enables us to reuse the various existing, highly optimised electrostatic libraries/codes without any specific modification. In addition, the various boundary conditions/corrections (*e.g.*, single or double periodicity[77, 103, 104] or the constant surface charge[105]) have been implemented in these libraries. They could be readily included in the corresponding $\mathbf{E}$ without the additional treatment. Moreover, our method can also be extended to solve other electrostatic constraints, *e.g.*, the ICC* method for the dielectric boundary[17, 18] or Qeq method[106], or to derive the differential interaction matrices of the electrostatic energy, such as the electric force or the electric virial tensor.

## *2. Preconditioned conjugate gradient (PCG)*

Even with our new method to calculate $\mathbf{E}$ with high efficiency, the determination of $\mathbf{q}^{\text{ele}}$ from Eq. (35) still needs some careful treatment. In Eq. (35), the product between $\mathbf{E}^{-1}$ and $\mathbf{b}$ to derive $\mathbf{q}^{\text{ele}}$ involves $N^2$ times of multiplication operations. This product needs to be taken in every MD simulation step, and the high computational cost (red curves in Fig. 7b) limits the application to an extensive system. Moreover, the quantum band structure would yield a non-linear electrochemical energy functional beyond the capability of MI techniques.[33] Thus, the matrix-*free*, iterative method, *i.e.*, CG, is widely used to solve Eq. (35) in CPM simulation, which scales quasilinearly (grey curve in Fig. 7b). In this section, we further accelerate the conventional CG method by solving it in a preconditioned manner as $\mathbf{PEq}^{\text{ele}} = \mathbf{Pb}$, where $\mathbf{P}$ is the (sparse) preconditioner.

Figure 2a presents the structure of $\mathbf{E}$ for 50 Gaussian charges in a cubic periodic cell with $L = 75.0$ Å and $\alpha_i = 1.2$ Å$^{-1}$. Figure 2b shows $\mathbf{E}^{-1}$ and the conversion into a sparse matrix $\mathbf{P}$ by removing the negligible (far-away) off-diagonal elements. The blue crosses in Fig. 2c indicate the removed (zero) entries in $\mathbf{P}$. In this case, *the* $r_{\text{cut}}$ is set to be 15 Å, and the number of neighbour atoms, including itself, is $N_{\text{cut}} = 21$. Note that the density of $\mathbf{P}$ controls the number of multiplication operations (cost) and the rate of convergence of PCG (effectiveness). For example, if all entries are set as $\mathbf{P} \equiv \mathbf{E}^{-1}$, the PCG becomes equivalent to MI and reaches full convergence in a single iteration. Thus, the optimal performance of PCG must be a good compromise.

Considering the neighbour list implementation in the modern MD simulation package, we sparsify $\mathbf{E}^{-1}$ based on the separation distance of two charges of $q_i$ and $q_j$ in real space ($r_{ij}$) to determine $\mathbf{P}$. We adopt two options: PCG-Full and PCG-Normal. In the PCG-Full method, $\mathbf{P}$ holds all entries of $e_{ij}^{\widehat{-1}}$, given that the two charges are within the same neighbour list. The PCG-Normal method further requires the separation distance $r_{ij}$ would be smaller than the preset $r_{\text{cut}}$. After reviewing the structure of $\mathbf{E}^{-1}$, we found that $e_{ij}^{\widehat{-1}}$ decayed exponentially with the separation distance. Thus, it might be possible to generate the high-quality $\mathbf{P}$ based on the *local* environment of the electrode atoms, like the nearsightedness theorem[107] in the density functional theory (DFT) study. We proposed the third method to construct $\mathbf{P}$ as a Divide-Conquer manner (PCG-DC). The following steps illustrate the PCG-DC scheme in Figure 2d-f,

1) The $\tau$th electrode atom is selected, and the local $N_{\text{cut}} \times N_{\text{cut}}$ interaction matrix $\mathbf{M}_\tau$ among all its neighbourhood within $r_{\text{cut}}$ is produced (Fig. 2d from the green square of Fig. 2a), where $N_{\text{cut}}$ is the number of the neighbourhood, including $\tau$ itself. Each entry $m_{ij}$ is determined by distance $r_{ij}$ and a screening factor $g$.

$$m_{ij} = \begin{cases} \sqrt{\dfrac{2}{\pi}}\alpha_i & i = j \\ \dfrac{\text{erfc}(g|r_{ij}|)}{r_{ij}} - \dfrac{\text{erfc}(g|r_{\text{cut}}|)}{r_{ij}} & i \neq j, \quad r_{ij} < r_{\text{cut}} \\ 0 & i \neq j, \quad r_{ij} \geq r_{\text{cut}} \end{cases} \quad (36)$$

2) The local matrix $\mathbf{M}_\tau$ is inverted in Fig. 2e. The column of $\mathbf{M}_\tau^{-1}$ that represents the interaction of the atom $\tau$ with its neighbourhoods is filled back into the $\tau$th column of $\mathbf{P}$ (red box in Fig. 2e&f). The entries in the $\tau$th column of $\mathbf{P}$ with $r_{\tau j} > r_{\text{cut}}$, *i.e.*, beyond the red rectangle in Fig. 2f, are set as zero. Note that the sequence (*i.e.*, column or row index) of the $\tau$th atom and its neighbourhoods in $\mathbf{M}_\tau$ might differ from that in the final $\mathbf{P}$.

3) Repeat steps (1)-(2) by $N$ times for all electrode atoms until all columns of $\mathbf{P}$ are filled with the same sparsity as in Fig. 2c.

Moreover, by comparing the structure between $\mathbf{P}$ and $\mathbf{E}^{-1}$, the performance can be further improved by replacing the zeroes in $\mathbf{P}$ with $p_{\text{gap}}$, defined as

$$p_{\text{gap}} = \dfrac{1}{n_{\text{zero}}}\left(\sum_{i,j} e_{ij}^{-1} - \sum_{i,j} p_{ij}\right) \quad (37)$$

Here, $p_{\text{gap}}$ is the average of the removed entries during the sparsification. The $n_{\text{zero}}$ is the number of zeros in $\mathbf{P}$. Note that the 1st term is the sum of $\mathbf{E}^{-1}$, which could be easily calculated in the iterative approaches (see details in Supplementary Note).

In our iterative PCG approaches, the relative change of the interaction energy between electrode and electrolyte, $\Delta E$, is employed as the stopping criterion,

$$\Delta E^k = \dfrac{\sum_i \Delta q_i^k b_i^{\text{sol}}}{\sum_i q_i^k b_i^{\text{sol}}} = \dfrac{\sum_i \Delta q_i^k (\sum_I q_I e_{iI})}{\sum_i q_i^k (\sum_I q_I e_{iI})} \quad (38)$$

where $\Delta q_i^k$ is change of the $i$th electrode charges at the $k$th iteration step, and $q_i^k$ is the determined charge. $\mathbf{b}^{\text{sol}}$ represents the potentials/interactions by the electrolyte molecules/ions. Like most iterative techniques (Fig. 6), $\Delta E^k$ decreases exponentially with the iteration step. At a large enough $k$, the summation of $\sum_i q_i b_i^{\text{sol}}$ becomes a reasonable estimate of the convergent results at $k \to \infty$.

Similarly, $\sum_i \Delta q_i^k b_i^{sol}$ at large $k$ is a good approximation of the overall residual. Thus, $\Delta E^k$ at the $k$th step is employed as the convergent criterion. Compared with the other criteria, such as the maximal change of relative charge, e.g., $\max|\Delta q_i^k/q_i^k|^{18}$, our criterion endows the residual of charges $\delta q_i^k = q_i^k - q_i^\infty$ with the different weight as $b_i^{sol}/\sum_i q_i b_i^{sol}$. The consistency of our PCG techniques with the MI one is also given in the following sections.

## C. Choice of $\alpha_i$

At present, the choice of $\alpha_i$ (the distribution width of the Gaussian charges on the electrode) is determined empirically in CPM. Several studies[24, 34, 108] have concluded that the $\alpha_i$ values could significantly change the polarizability of conducting surfaces and influence the electric double layer properties at the interface[109]. It is valuable to develop an approach to determine $\alpha_i$ based on physical principles or models.

Considering that the CPM method focuses on predicting ion storage and dynamic properties at charged surfaces, it is reasonable to propose that the adopted $\alpha_i$ in the discrete atomistic model should reproduce similar properties/performances of the charged electrode/ionic system (e.g., capacitance) as the continuum limit.[10,11] Here we investigated the capacitance and the electric potential profiles in an atomic graphene capacitor model with two uniformly, oppositely charged graphene sheets (inset in Fig. 3). Within a 51.12 × 54.11 × 50 Å periodic cell, two graphene sheets are located at $z = \pm 12.5$ Å. An Ewald summation is then used to perform the electrostatic calculation for the point charges on the graphene atoms. Figure 3 shows the potential and electric field profiles using a small testing charge. The electrical potential results are along the $z$-direction (passing through the centre of a benzene ring). The red-dash line is the result of analytic continuum limit models.

The atomistic and the continuum limit models show identical curves except in the regions close to the sheet surface (< 2.5 Å). In this region, the electrode atomic charges cannot be regarded as a uniform distribution on an ideal metallic surface (continuum limit), resulting in a smooth transition of the electric field and a lower electrode potential. The grey horizontal bars represent the corresponding potentials of the graphene atoms. The $U^{GRA}$ in the inset of Fig. 2 is the electric potential discrepancy between the discrete atomic model and the continuum limiting. It is inversely proportional to the lattice constant or the nearest neighbour distance ($r_{1NN}$).

If the electrode atom charges (graphene) were replaced by Gaussian charges, the electrode potentials would be elevated by their self-energy, which is proportional to the inverse width of $\alpha_i$. For graphene with carbon bond length $r_{1NN} = 1.42$ Å, $\alpha_i^0 = 2.933$ Å$^{-1}$ would precisely compensate the underestimated electrode potential, as $U^{GRA} = 2/\sqrt{\pi}\alpha_{ii}^0 = \sqrt{2/\pi}\alpha_i^0$. The dashed cyan curve is the corresponding Gaussian potential of Eq. (5) using the decided $\alpha_i^0$. Despite the asymmetric swelling of the electric field on the graphene surface, the $\alpha_i^0$ minimises the potential difference between the two models.

Attributed to the Graphene lattice symmetry, the normalization of $U^{GRA}$ with $r_{1NN}$ is equal to a lattice constant, i.e., the Madelung constant of 2D graphene-like Wigner crystal $U^{GRA}/r_{1NN} = -3.32302166466011$. Thus, a series of of $\alpha_i^0$ or $U^{LATT}$ for the graphene and other 2D lattices[39] are analytically determined and summarized in Table 1.

For the graphene electrode system, it is worth noting that the determined $\alpha_i^0$ is much smaller than the electrode interatomic distance ($\sim$1.42 Å) and the electrode-electrolyte distance ($>$ 2.0 Å). The short-range screening of the Gaussian charges becomes negligible ($< 3.2 \times 10^{-5}$ in the last column of Table 1). Therefore, we can simulate the conducting graphene electrode surface using the classic point-charge with an additional non-singular self-energy $U^{GRA} = \sqrt{2/\pi}\alpha_i^0$. Note that Vatamanua et al. have taken a similar implementation empirically but did not provide a solid analysis.[25] Nakano and Sato reported that removing the screening factor, i.e., $\text{erf}(x)$, only negligible changes the structure profiles.[34] In the following study, we will use such point-charge simplification for graphene electrode systems, which will simplify the real-space calculation.

From the perspective of quantum mechanics[33, 34], the original SR-CPM framework is a simplified, second-order expansion of the total electrochemical energy. The electronegativity of the material $\chi^0$ is omitted for the symmetric electrode materials. Meanwhile, the complex interaction kernel of $J_{ij}^0$ between the electrode charges and the chemical hardness $J_{ii}^0$ or Hubbard-U $U_i^0$ are approximated by the interactions of the Gaussian charge with $J_{ii}^0 = \varphi^G(\mathbf{r}; \alpha_{ij})$ and $U_i^0 = \sqrt{2/\pi}\alpha_i^0$. A more realistic model would decouple both terms. It is worth noting that our $\alpha_i^0$ is proposed to describe an ideal metallic surface rather than the real electrode, with a significantly different electronic structure. Concerning the latter, e.g., by using the experimental $U_i^0$ (Ref. [34]) or considering Thomas-Fermi screening[35], the electrode characteristics could be altered.

To verify the proposed $\alpha_i^0$, we studied the interaction of a point charge ($q = +1$ e) to a polarizable conductor surface (graphene). The simulation cell of 102.1 × 100.7 × 204.2 Å was symmetrically set up to include two graphene sheets (3936 atoms at $z = \pm 5$ Å) and two point-charges in $z$-directions. Using our MI CPM techniques, both graphene sheets were grounded ($\chi_i = 0$ V) under the electroneutrality constraint. The point charges were moved along the $z$-direction with a constant step size. Negative surface charges are induced with the point charge approaching the graphene surface. For an ideal grounded metal, the surface charge is equivalent to an image charge of $q = -1$ e on the opposite side. Considering the limited dimension of the graphene sheets under the PBCs, the point charges and periodic images form a 2D periodic pattern in the $x$- and $y$-directions. The accurate interaction energy of such a 2D charge pattern could be given by the MMM2D approach and compared with our CPM work.

Figure 5 shows the relative error of the point-charge/graphene interaction energy and force with different $\alpha_i$. $\Delta z$ is the distance between the point charge and graphene. Each data point was averaged over 4000 individual samples obtained by randomly displacing the point charges in the $xy$-plane. The force and energy were calculated via a highly accurate Ewald summation (below $10^{-8}$). As expected, our CPM simulation reproduced the behaviour of the ideal metallic surface with an induced image charge.[110] Both the magnitudes and the spreads of error grow as the point charge approaching graphene surface. It is because the distance between the discrete charges on graphene becomes comparable with $\Delta z$. The red-dash lines in Fig. 5 represent the result from the PCG-DC approach with the convergence criterion of $10^{-4}$, which is similar to the black curves using the MI method.

The error was minimal at the determined value $\alpha_i^0 = 2.93295072600296$ Å$^{-1}$ (< 0.5%). Some previous work[34, 108] reported the decrease of $\alpha_i^0$ enhanced electrostatic interaction and vice versa, which is consistent with our results. The abrupt change at $\Delta z = 10$ Å is due to the cutoff error in real space. We performed similar studies for the other typical 2D lattices, *i.e.*, square, hexagonal, and Kagome lattices, in Fig S2-4, respectively. Minimal errors do occur at these determined $\alpha_i^0$ values in Table 1. Besides the 2D structure, our results could be applicable to electrodes with a 3D lattice structure. For example, $\alpha_i = 1.979$ Å$^{-1}$ was ever used to describe Pt (111) surface with the lattice constant of $a^{\text{FCC}} = 3.9$ Å.[11] If the outermost layer of Pt is treated as a 2D hexagonal lattice with $a = 2.758$ Å, the optimised $\alpha_i^0$ would be $\alpha_i^0 = \sqrt{2/\pi}\frac{U^{HEX}}{a} = 1.915$ Å$^{-1}$. It is close to the employed value of 1.979 Å$^{-1}$ in previous work. Another example is the employed $\alpha_i = 1.787$ Å$^{-1}$ at the Pt (100) surface with $a^{FCC} = 3.92$ Å.[10] It is also close to the determined $\alpha_i^0 = 1.764$ Å$^{-1}$ for a 2D square lattice on the (100) surface.

It is worth noting that removing the electroneutrality constraint drastically deteriorates the result,[100], especially at a significant distance, e.g., nearly 50% relative error at $\Delta z = 18$ Å. In this case, the induced surface charge ($< 1$ e) decreased, and electrostatic interaction was underestimated. As a result, the global electroneutrality constraint was always used in the following CPM research that simulates the systems under a constant potential difference.

# IV Computation efficiency & Verification

## A. Convergence rate of PCG

This section studies the convergence rate of the iterative PCG approaches for a typical graphene electrochemical cell with 2 nm slit size (Fig. 4). The simulation cell consisted of two ~81.2 × 89.5 Å graphene sheets ($N = 5544$), filled with 1 M NaCl electrolyte of 3906 water molecules and 72 pairs of ions. The water molecules and the ions were in equilibrium state at 300 K. A voltage difference of 1 V between two graphene sheets was applied in our CPM simulations. The different potential constraints $\chi_i$ applied on the cathode and anode break periodicity in the $z$-direction. Only in-plane 2D periodicity is used in our electrochemical simulations. The Yeh-Ballenger boundary correction term (BC) [103, 111] is employed in the z-direction. It is incorporated into the conventional 3D Ewald summation with the additional BC terms in each $e_{ij}$ as follows,

$$e_{ij}^{BC} = -\frac{2\pi}{V}\left[(z_i - z_j)^2 + \frac{L_z^2}{6}\right] \quad (39)$$

where $z_i$ denotes the $z$-position of the $i$th charge, and $L_z$ is the dimension of the cell in the nonperiodic $z$-direction. The 1$^{st}$ term counterbalances the contribution of the net dipole momentum.[77] The position-independent 2$^{nd}$ term is the slab non-neutral correction $Q \neq 0$ by adding two opposite surface charges on the boundaries.[103]

Figure 6 shows the convergence rates for different iterative techniques. In Fig. 6a, performance of the iterative approaches follow the order: PCG-Full > PCG-Normal: $r_{cut} = 8$ Å > $r_{cut} = 4$ Å > $r_{cut} = 2$ Å > $r_{cut} = 1$ Å where $r_{cut}$ is the *cutoff* for the sparse **P** in PCG-Normal. The PCG method exhibited superiority over the traditional CG in most cases (grey curves in Fig. 6), except for $r_{cut} = 1$ case with an inferior **P**. This could be attributed to the adopted cutoff $r_{cut} = 1$ Å in the last case is smaller than the distance to their nearest neighbours. The free **P** shift with $p_{gap} = 0$ (half-filled) shows the marginal improvement in the effectiveness, especially for the PCG-Full method.

Figure 6b summarizes performance of the PCG-DC techniques with different $g$ at $r_{cut} = 8$ Å. The value of $g$ controls the convergence rate by changing the matrix structure (*i.e.*, local **M** and **P**). The optimal value $g = 0.083$ Å$^{-1}$ is obtained through a trial-and-error process. Compared to the PCG-Normal approach, which includes matrix calculation and inversion, the computation of **P** in PCG-DC is highly parallel and takes only a few seconds (< 10 s). The exponential decay of the entries in $\mathbf{E}^{-1}$ is responsible for the efficacy of the DC method. Such decay could arise from the charges' strong self-interaction, resulting in a diagonally dominating structure in **E** as $e_{ii} = \sqrt{2/\pi}\alpha_i \gg e_{ij}$. The optimal $g$ value could be determined through a one-time trial-and-error process for the electrode with the different lattice structures.

## B. Computational efficiency

Similar to previous section, a series of graphene electrochemical cells with 2 nm slit size were simulated to test the scaling laws of the different CPM methods. The dimensions of the simulation cell, *i.e.*, the lateral size of the graphene sheet, increased from ~30 to 3430 Å. The number of electrode atoms, water, and ions increased proportionally from 616, 434, and 8 to several million. Our implementation with the P3M solver (P3MDC) is compared with that of the previous Wang's implementation[4] following the seminal CPM work using a Gaussian Ewald summation (EW3DC). In the P3MDC, the P3M calculation was taken with $\eta = 0.32508$ Å$^{-1}$ and the ratio of the FFT grid numbers to the dimension of the simulation cell was fixed, *e.g.*, 18 × 18 × 30 for the smallest cell of 27.1 × 29.8 × 20.0 Å.

We implemented our CPM code in the LAMMPS MD simulation package (**conp/ME**), abbreviated from the *Constant Potential Simulation Method for Metal Electrode*, which is applicable to not only the conducting graphene sheet but also to the arbitrary metallic surface. The whole LAMMPS package and our CPM code were compiled by the Intel C++ Compiler of v19.0.4.243 and linked with the associated Intel Math Kernel Library (MKL) numeric library (20190004) and Intel Message Passing Interface (MPI) Library (20190430). Our tests were performed using 24 cores (AMD EPYC 7742 64-Core Processor) on a Linux workstation.

Figure 7a illustrates the time cost of **E** calculation. In EW3DC, the elements of **E** are calculated pairwisely by $N^2$ times, using an internal Gaussian Ewald summation $\mathcal{O}(N^2)$ that resulted in a total biquadrate time of $\mathcal{O}(N^4)$ with prohibitive memory cost. In contrast, our P3MDC exhibits a quasi-quadratic scaling $\mathcal{O}(N^2 \log N)$ by only taking $N$ times of the linearithmic P3M calculation. We also implemented the intermediate matrix approach (orange curve)[91] with a similar $\mathcal{O}(N^2)$ scaling in the LAMMPS simulation package. It exhibits slightly better performance than our P3MDC method (about 1.5~1.6 times faster in **E** calculation). However, the prohibitive memory cost prevents it from being used in extensive systems ($N > 10^4$). Note that when $N > 10^4$, the inverse of **E**, using the serial LAPACK library as Wang's CPM implementation, takes a longer time than the calculation of **E**. Therefore, we employed the distributed parallel ScaLAPACK library to speed up the calculation, which accelerated the matrix inversion by approximately 14.6 times (Fig. S6).

Figure 7b compares the elapsed time of a single MD step using the EW3DC and P3MDC methods. The elapsed time is an average of 1000 steps at the equilibrium (300 K) using the MI approaches. The EW3DC technique, like the previous studies[91], has poor scaling performance compared to P3MDC. The total cost to derive $\mathbf{q}^{ele}$ in the MI approach is further divided into three components in Fig. 7c: the matrix-vector multiplication, the pairwise real-space calculation, and the $k$-space calculations. The latter two calculations compute the potential on the electrode from the electrolyte molecules and ions, $\mathbf{b}^{sol}$. Due to reusing the existing P3M code in the P3MDC, both real- and $k$-space calculations show a quasilinear scaling, in contrast to the $\mathcal{O}(N^{1.5})$ or $\mathcal{O}(N^2)$ scaling in the EW3DC method.

Figure 7b also shows the time cost ratio (blue curve) between the P3MDC and the FCM using the same configuration. The cost ratio increases considerably from ~1.93 at $N = 616$ to ~6.51 at $N = 7.4 \times 10^4$. It is due to the quadratic scaling of the matrix-vector multiplication in the MI approach

(Fig. 7c). As a result, iterative approaches should be a promising candidate to simulate the metallic electrode for a sizeable system ($N > 10^4$).

Figure 7d shows the elapsed time of a single MD step using the CG or PCG-DC approaches. The convergent criterion is $10^{-8}$ for both methods. It indicates the superior performance of the iterative approaches over the MI approach for the sizable system $N > 10^4$. The PCG-DC method exhibits 1.5~2.0 times better efficiency than the conventional CG approach. By decreasing the criterion to $10^{-4}$, the cost is generally reduced by about 20~40% (Fig. S5). The influence of the iterative techniques on the structural and dynamical properties, especially by the convergent criterion, should be thoroughly reviewed for practical application, which would be taken in the future.

To demonstrate the scalability of our CPM implementation, we enlarged the size of the simulation cell up to 3111.9 × 3429.3 × 20 Å. Our PCG-DC approaches can efficiently handle the extra-large system containing over 8.1 million electrode atoms, 5.7 million water molecules, and 0.1 million pairs of NaCl ions. The total number of atoms is about 12.2 million in this case, and the FFT grids are 2160 × 2160 × 30. Note that the PCG-DC technique generates $\mathbf{P}$ highly parallelly, taking only ~79 seconds. Nevertheless, the scale-up of the system size deteriorated the efficiency of the PCG-DC CPM simulations compared with the FCM simulations. The blue curve in Fig. 7d shows the time cost ratio between the PCG-DC and FCM methods. It grows from a minimal ~2.0 at $N = 10^4$ to a maximum of ~7.5 at $N = 8.1 \times 10^6$. Because the dimension of $\mathbf{E}^{-1}$ expands quadratically as the size increases, whereas the nonzero of $\mathbf{P}$ grows linearly, more and more electrode atoms will be located beyond the pairwise cutoff ($r_{\text{cut}}$) with increasing system size, which reduces the efficiency of the preconditioning techniques.

## C. Verification

This section verifies our CPM implementation (P3MDC) in the LAMMPS MD simulation package by comparing it with the previous Wang's implementation (EW3DC). The simulation cell consisted of two ~81.2 × 89.5 Å graphene sheets ($N = 5544$), filled by 1 M NaCl electrolyte of 3906 water molecules and 72 pairs of ions (Fig. 4). We adopted the TIP3P water molecule.[112] The TIP3P water molecule is fixed through the shake algorithm[113], and Lennard-Jones-12-6 (LJ) potential coefficients for aqueous ions were taken from Li's work[114], calibrated with hydration free energy. A Nose-Hoover thermostat[115] was adopted to fix the temperature at 300 K.

In P3MDC, the electrostatic interaction is accounted for by the P3M method with an accuracy of $10^{-4}$. While EW3DC computed **b** and **E** using an internal Gaussian Ewald summation, that reduced the computation efficiency by one order of magnitude. The widths of the Gaussian charges in both CPM simulations were set to be $\alpha_i^0 = 2.93295072600296$ Å$^{-1}$ for the graphene lattice. The MI technique was employed. The crucial electroneutrality constraint was employed in P3MDC but absent in EW3DC. The point-charge simplification was also used to accelerate the P3MDC calculation. Both methods showed a similar average electrode charge of 0.004500 vs 0.004411 e per atom under the 1 V potential difference. Besides, the results of our PCG-DC method are given in Fig. S7 with a similar structure as the full convergent MI method.

Figure 8 shows the average distribution of ions and water and the potential profiles along the z-direction. For clarity, the profiles of water molecules are plotted on the lower negative axis of Fig. 8a. The distributions in EW3DC and P3MDC were consistent with each other, where the counterions are absorbed onto the charged surface, in contrast to the depletion of the co-ions. The oppositely charged graphene sheets lead to non-symmetric water distribution, *e.g.*, the different oxygen peaks on two sides, ~3.18 vs 3.49, and the different extensions of hydrogen distribution. The tilt of the O-H bond is more significant on the anode surface (negative charged).

It is worth noting that there are minor differences in ion distributions between P3MDC and EW3D. For example, the heights of the right Cl$^-$ peaks are 4.32 and 4.51 in the P3MDC and EW3DC results, respectively. Similar results can be observed on the left Na$^+$ peaks with 2.54 and 2.35, respectively. The slight discrepancy might be attributed to the absence of the electroneutrality constraint in EW3DC.

The potential profiles in Fig. 8b were calculated by integrating the 1D Poisson's equation, $\partial^2 \phi / \partial z^2 = -4\pi\rho(z)$ where the factor $4\pi$ is added from the Gaussian units. The overlapped ions and water distributions resulted in nearly identical profiles. Similar to the previous work[96], the total potential (red) in Fig. 8b was divided into two parts: the contribution from the surface charge and the mobile ions (blue) and the remaining from the dielectric response of the water molecules (orange). The former potential was flattened by the presence of ions that lowered the voltage difference (from ~6.22 V in vacuum to ~3.10 V). The linear region on both sides indicated the potential growth of the surface charges on graphene, which is derived from average electrode charges. Then, the water molecules as the dielectric medium produced an opposite potential curve with a drop of −2.11 V. The combination of two potentials finally resulted in the ~1 V voltage between the two sides in all three studies (red curve). The negligible voltage difference (~0.005 V) originated from the slight discrepancy in the ion profiles in Fig. 8a. It is attributed to the breakdown of electroneutrality under the Yeh-Berkowitz correction [91] that leads to the slightly asymmetric average electrode charges (0.004411 vs. 0.004492 e) in the EW3DC.

Figure 8 describes the distributions and potentials by FCM as well. The simulation setup in FCM is the same as that of P3MDC except for the invariant surface charge on the graphene. By charging the graphene at 0.0045 e per atom (average electrode charge in P3MDC), the profiles in FCM are consistent with those of P3MDC, except for stronger ion adsorption in P3MDC. For example, the density of the Cl⁻ peak increased from 4.02 to 4.32. The improved adsorption capability is related to the image-charge effects.[3, 116] Nevertheless, the strong hydration shell of the cation $Na^+$ weakens the image-charge adsorption with the smaller relative density increased from 2.43 to 2.54. The overlapped potential profile in Fig. 8b suggests again that the low voltage EDLs structure in the FCM would be similar to that in CPM.[3, 4]

## V. Conclusion

In this work, we carefully reviewed the mathematical background of the Ewald summation and the physical framework of SR-CPM. We reformulated the Ewald summation for point and Gaussian charges into a unified framework. As a result, the conventional meshed-based Ewald summation methods/codes could be employed (P3M method in our study). The electroneutrality constraint plays a crucial role in determining the polarizability of the simulated metal surface. Thus, a general post-treatment step has been proposed to neutralise the system, which is applicable for both matrix and iterative methods and make it feasible to study the half-cell configuration with the monopolar electrolyte ($Q^{sol} \neq 0$). Besides, a series of the analytic $\alpha_i^0$ for 2D lattice structures were derived from the Madelung constant of the 2D Wigner crystal. The choice of $\alpha_i$ determines the spread of the Gaussian charge used in CPM and its self-energy. Our standard $\alpha_i^0$ excellently resembles the behaviour of the ideal metal surface and compensates for the potential gap between the discrete atomic model and the continuum limit $U^{LATT} = \sqrt{2/\pi}\alpha_i^0$.

This work proposes an improved method to calculate **E**, using $N$-times of mesh Ewald summation (P3M), giving a scaling of $\mathcal{O}(N^2 \log N)$, which becomes comparable with the recently developed intermediate matrix approach $\mathcal{O}(N^2)$. Our method is a general and straightforward approach. Therefore, it is favourable to reuse the various existing highly optimised electrostatic libraries/codes without specific modifications and to include the various boundary conditions considered in these libraries. Indeed, our method can also be applied to other electrostatic constraints or deriving high-order interaction matrices.

Due to the unsatisfactory performance of the MI approach in sizable system ($N > 10^4$), iterative techniques, *e.g.* CG, is another widely employed method in CPM. This work remarkably enhanced its performance by 1.5~2.0 times using the preconditioning technique (PCG). A highly parallel Divide-Conquer process (PCG-DC) is proposed to generate the optimised preconditioner **P**. Thus, the PCG-DC technique is feasible for the extensive system with an approximately doubled cost to maintain the potential constraint within $N < 2.4 \times 10^5$. The scalability of our PCG-DC implementation has been demonstrated on the 2 nm graphene electrochemical cell with over 24.2 million simulation atoms (8.1 million electrode atoms).

We implemented our new approaches (computing **E** using the existing optimised electrostatic code and the preconditioned CG methods) into a constant potential simulation module, **Conp/GA**, in the LAMMPS MD simulation package. Our in-depth analysis of the CPM method and the optimization of the numeric implementation code would help the community better understand the structure and dynamics of EDLs on the metallic nanochannel and further develop the CPM to simulate the more complex nanochannel with minimal error and maximum efficiency.

## SUPPLEMENTARY MATERIAL

The **supplementary material** contains the additional derivations involving the interactions between the Gaussian charges and the Fourier transform of the Ewald potential and the iterative method to obtain **c**, as well as the additional discussion on $\widetilde{\mathbf{Q}}$ in MI, Ewald summation for the Gaussian charges, $\alpha_i^0$ for 2D square or hexagonal lattice, the performance of CG and PCG methods, and the profile of PCG techniques in the graphene electrochemical cell.

## ACKNOWLEDGMENTS

G.J. acknowledges the support of the National Natural Science Foundation of China (grant no. 21905215) and the resources from High-Performance Computing Center of Wuhan University of Science and Technology. J.Z.L. would like to thank the Australian Research Council for their support (grant no. DP210103888).

## DATA AVAILABILITY

The source code of the **Conp/ME** package and the graphene electrochemical cell is available at https://github.com/kylincaster/LAMMPS_conp-ME

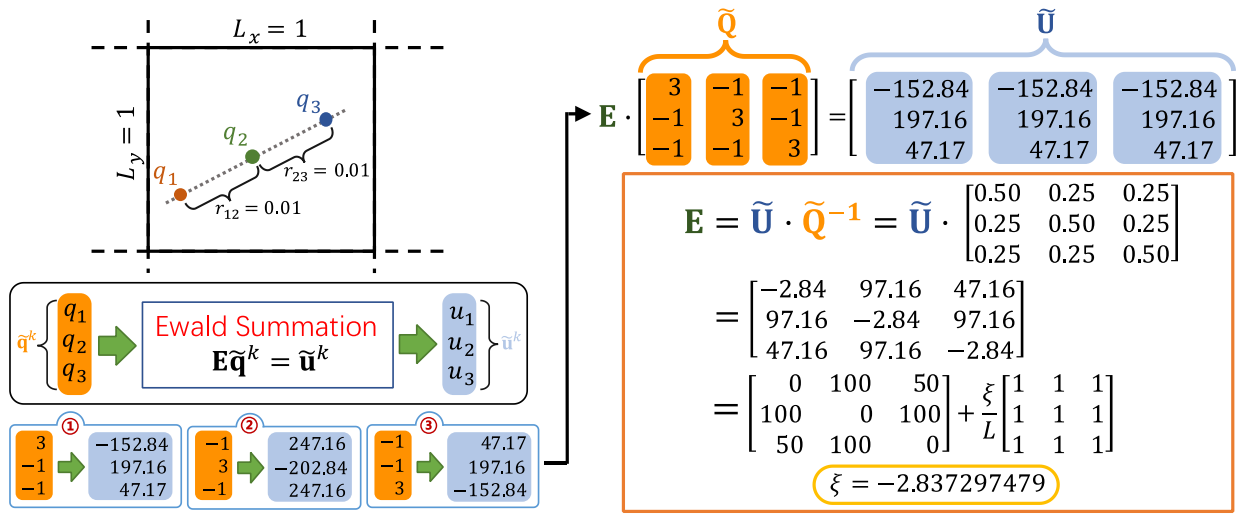

Figure 1: An illustration of our method to calculate the Ewald interaction matrix ($\mathbf{E}$) in a cubic cell of $L = 1$. Three point-charges are aligned in line with a spacing of $0.01L$. We take the Ewald summation to solve the potentials on the three charges $\widetilde{\mathbf{u}}$ by different charge configurations $\widetilde{\mathbf{q}}$ (the orange regions in the ①, ② and ③ boxes). The $3 \times 3$ symmetric matrixes $\widetilde{\mathbf{Q}}$ and $\widetilde{\mathbf{U}}$ are assembled (orange and blue regions) with the relation of $\mathbf{E}\widetilde{\mathbf{Q}} = \widetilde{\mathbf{U}}$. Note that $\xi$ inside $\mathbf{E}$ is the Madelung constant of an SC (Wigner) crystal.

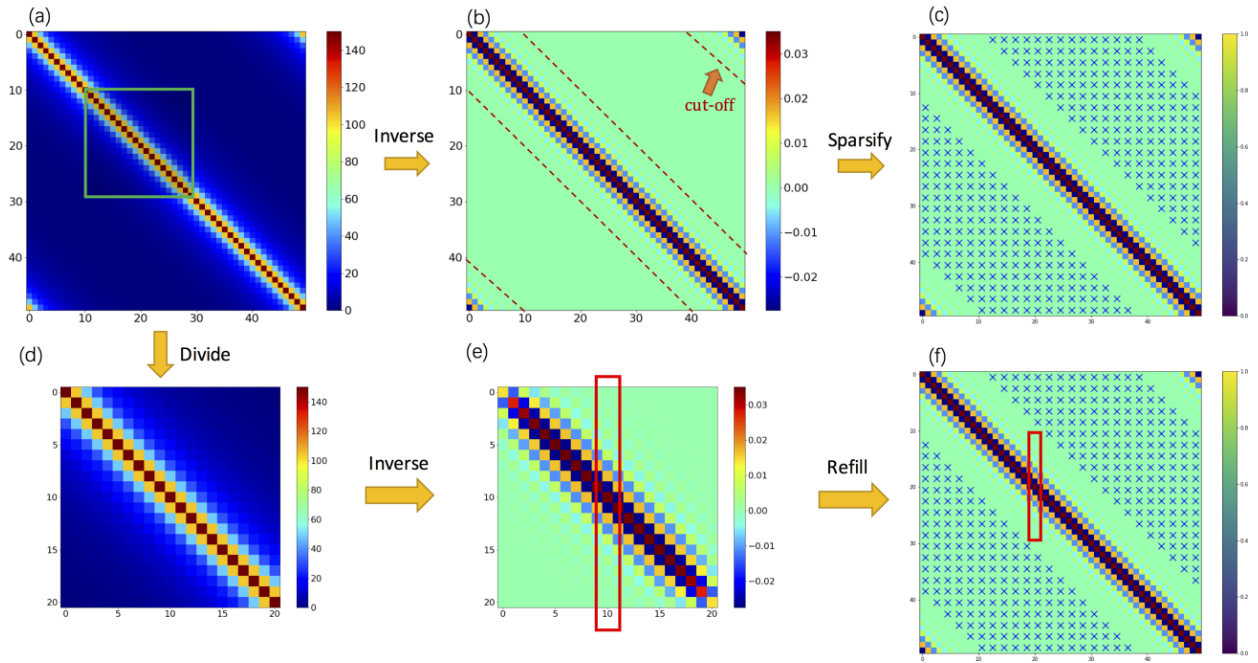

Figure 2: An illustration of our method to determine **P** in PCG for a 50-atom chain with $L = 75.0$ Å and $\alpha_i = 1.2$ Å$^{-1}$. The uppers (a-c) illustrate the PCG-Normal/Full approaches by directly sparsifying $\mathbf{E}^{-1}$ along with the real space distance. The lowers (d-f) describe the DC approach (PCG-DC). (d) is the localized $\mathbf{M}_{20}$ for the 20th electrode atom extracted from the green box in (a). (e) is $\mathbf{M}_{20}^{-1}$ inverted from (d), and the interaction column (red box) with its neighbourhoods is refilled back into **P**, as marked by the red box on the 20$^{\text{th}}$ column in (f).

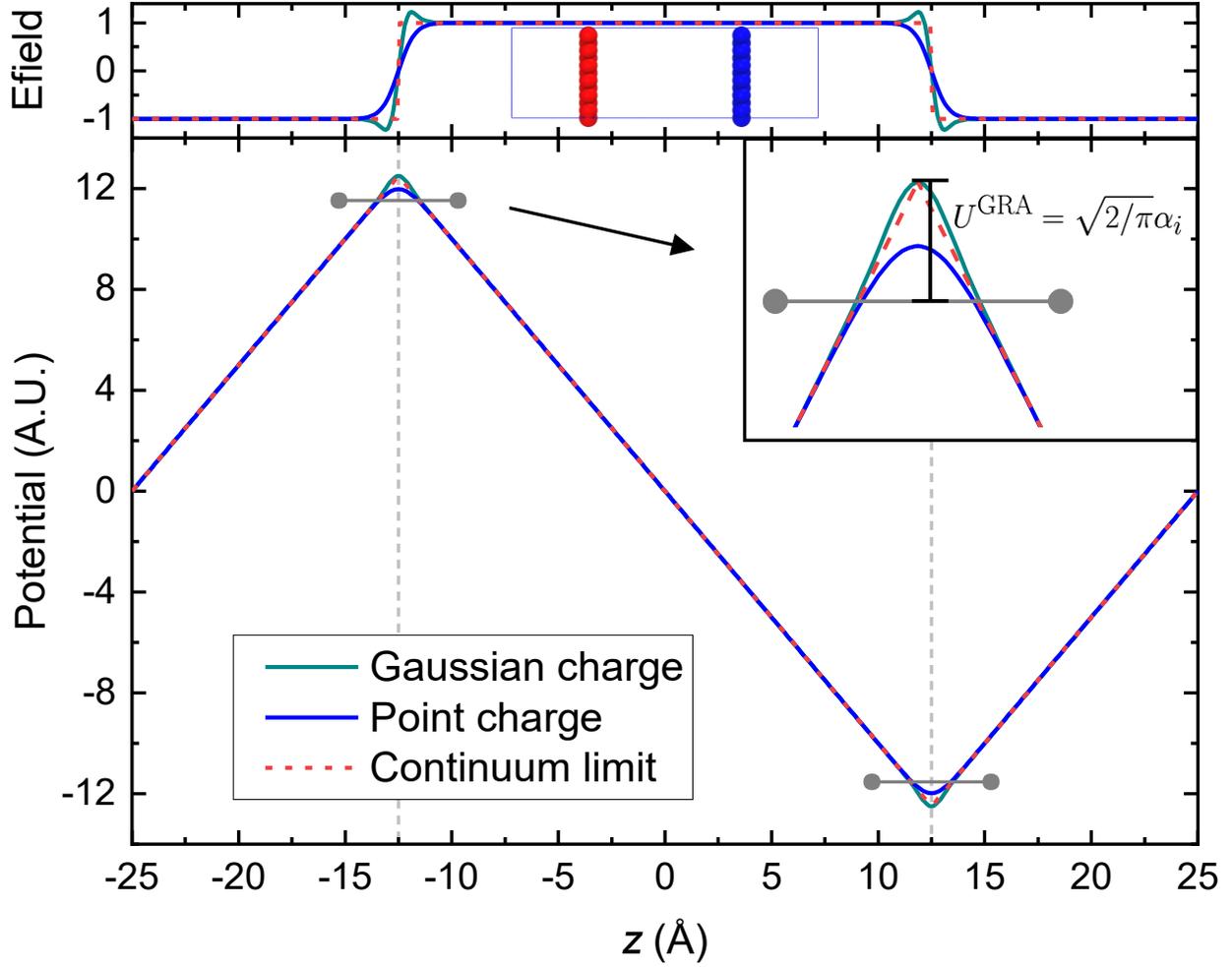

Figure 3: The distribution of electric field (upper) and potential (down) in a periodic graphene capacitor model. The solid cyan and blue curves show the profiles from the discrete atomistic model by the Gaussian ($\alpha_i^0 = 2.930$ Å$^{-1}$) and the point charges on the graphene atoms, respectively. The red-dash lines give the result of the corresponding continuum limiting. Two vertical lines mark the position of two graphene sheets ($z = \pm 12.5$ Å) of oppositely charged, and the horizontal grey bars indicate the electrode potential on graphene atoms. $U^{\mathrm{GRA}}$ in the inset shows the gap in the electrode potential between the atomistic point-charge and the continuum limiting models.

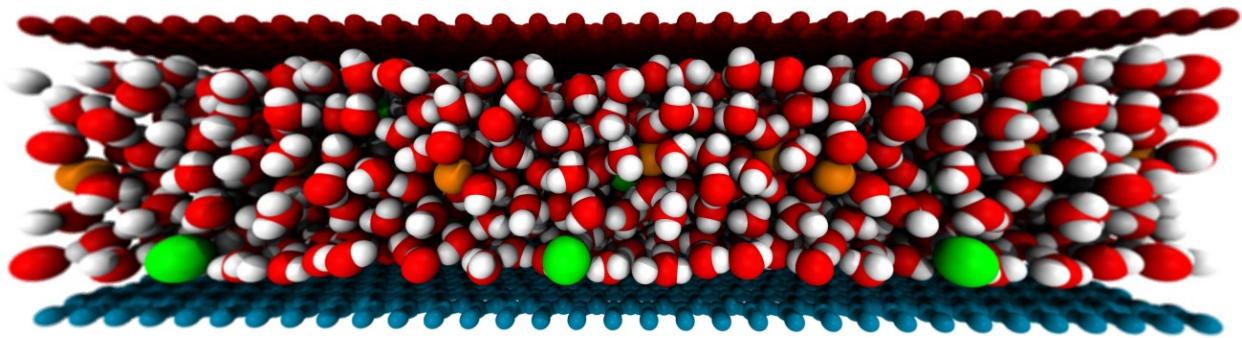

Figure 4: Schematics of MD simulation setup in which two ~81.2 × 89.5 Å graphene electrodes ($N = 5544$) sandwich 1 M NaCl electrolyte of 3906 water molecules and 72 pairs of cations and anions in a 2 nm nanoslit.

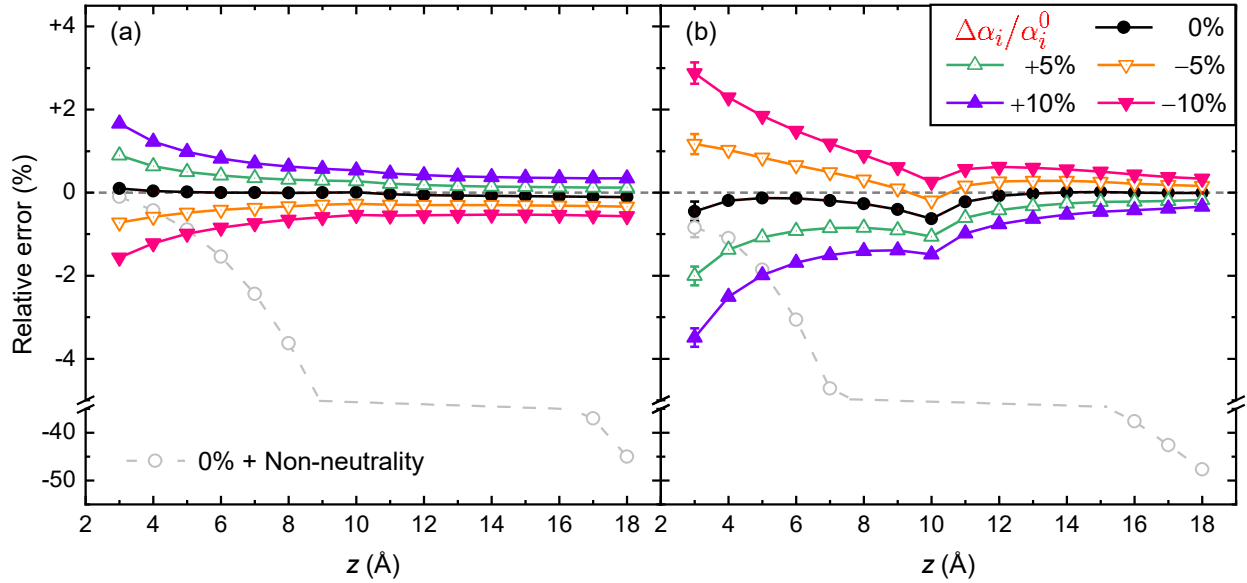

Figure 5: The mean and spread of the relative errors of energy (a) and force in the $z$-direction (b) versus the distance $\Delta z$ of the point charges to the graphene surface. The percentages on the right-axis describe the magnitude of the deviation ($\Delta\alpha_i/\alpha_i^0$) from the optimal $\alpha_i^0 = 2.9330$ Å$^{-1}$ for the graphene lattice. The error bars indicate the spread of 4000 samples. The open circles illustrate the non-neutral condition, and the red-dash lines are the results using a PCG-DC approach.

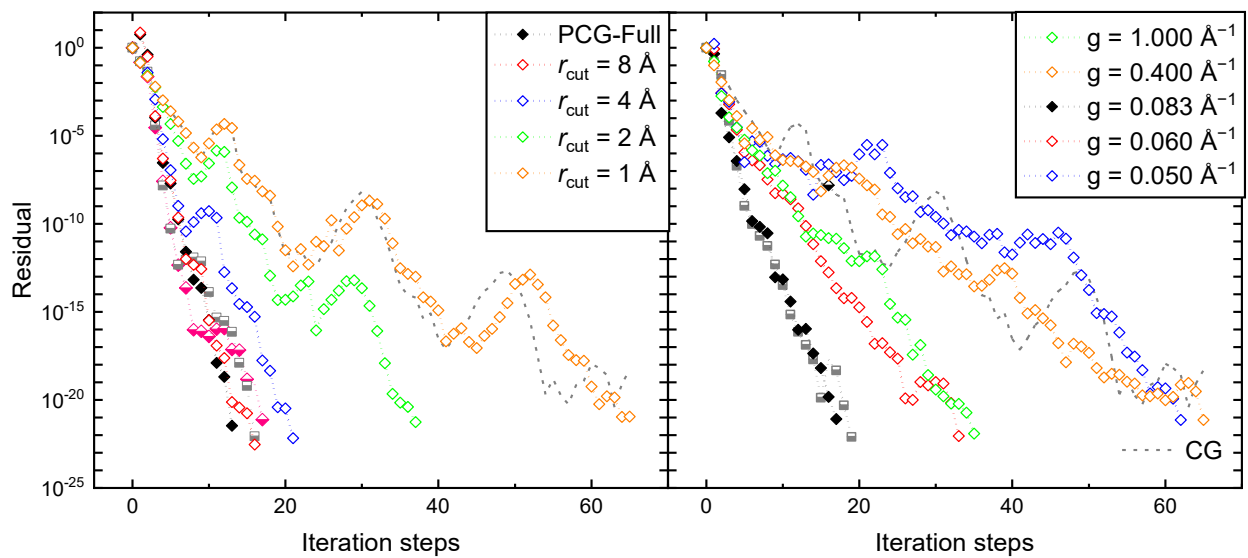

Figure 6: The convergence rate for different iterative approaches for a 2-nm graphene electrochemical cell. (a) The open and solid diamonds illustrate the results via the PCG-Full and PCG-Normal methods with the different cutoffs $r_{\text{cut}}$, respectively. The half-filled squares are the results of PCG-Normal with $r_{\text{cut}} = 8$ Å but free of **P** shift, *i.e.*, $p_{\text{shift}} = 0$. (b) PCG-DC method with different $g$. The half-filled squares indicate the results by PCG-DC with $g = 0.083$ Å$^{-1}$ and $p_{\text{shift}} = 0$. Note that the grey dashed is the performance of the CG method.

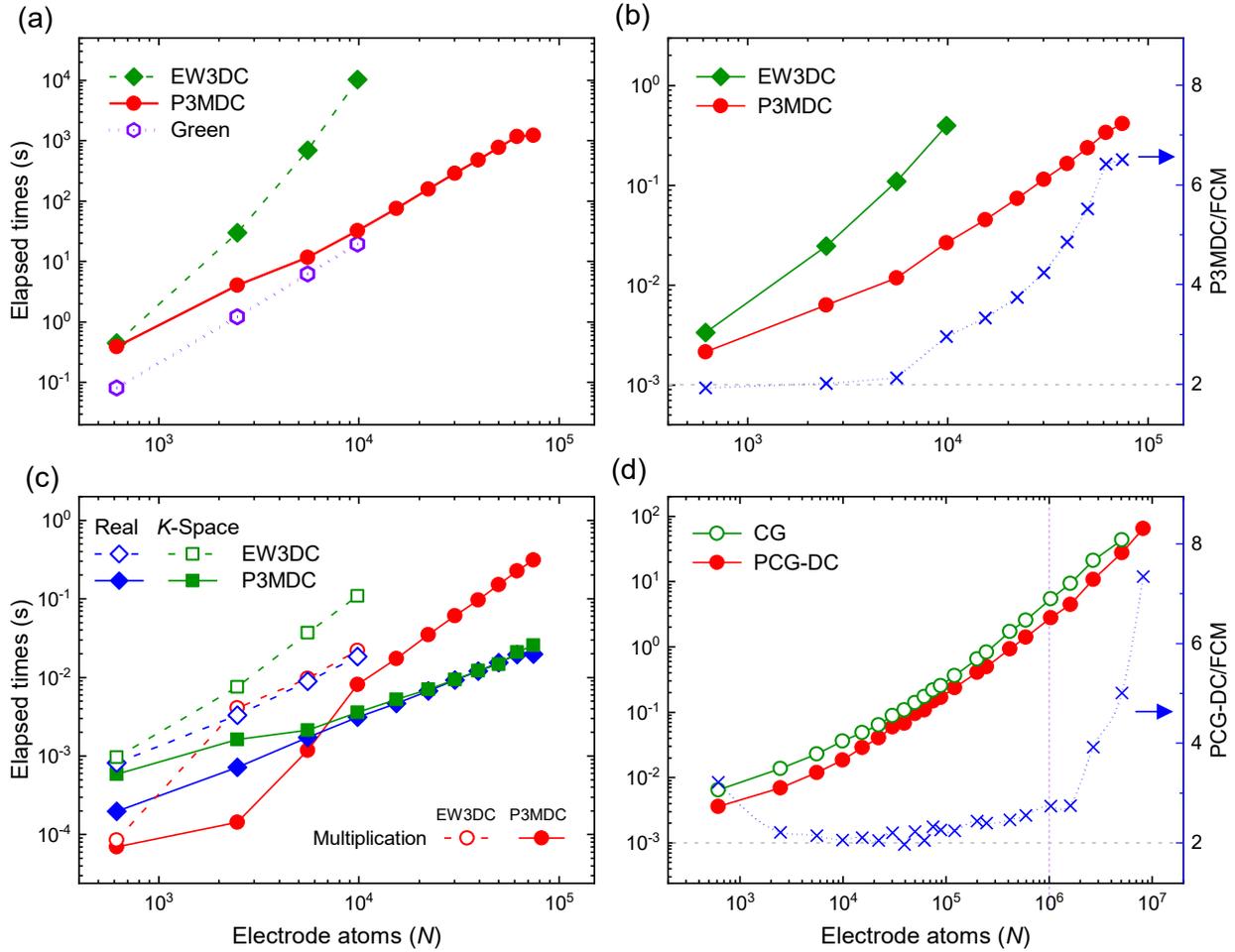

Figure 7: The computational efficiency of our CPM implementation versus the number of electrode atoms, $N$. (a) One-time **E** calculation cost in the EW3DC (green diamond) and P3MDC (red circle) methods. The purple hexagons illustrate the cost of using the intermediate matrix method.[91] (b) The single-step MD cost of the EW3DC (green diamond) and P3MDC (red circle). The relative cost to FCM simulation is also shown (blue cross). (c) The detailed cost components of the matrix-vector multiplication (red), real space (blue) and $k$-space (green) calculations in the MI approaches. (d) The single-step MD cost in the CG (open green), PCG-DC (solid red). A comparison with the FCM is also shown (blue cross). The vertical dashed line indicates the simulation of 1 million electrode atoms.

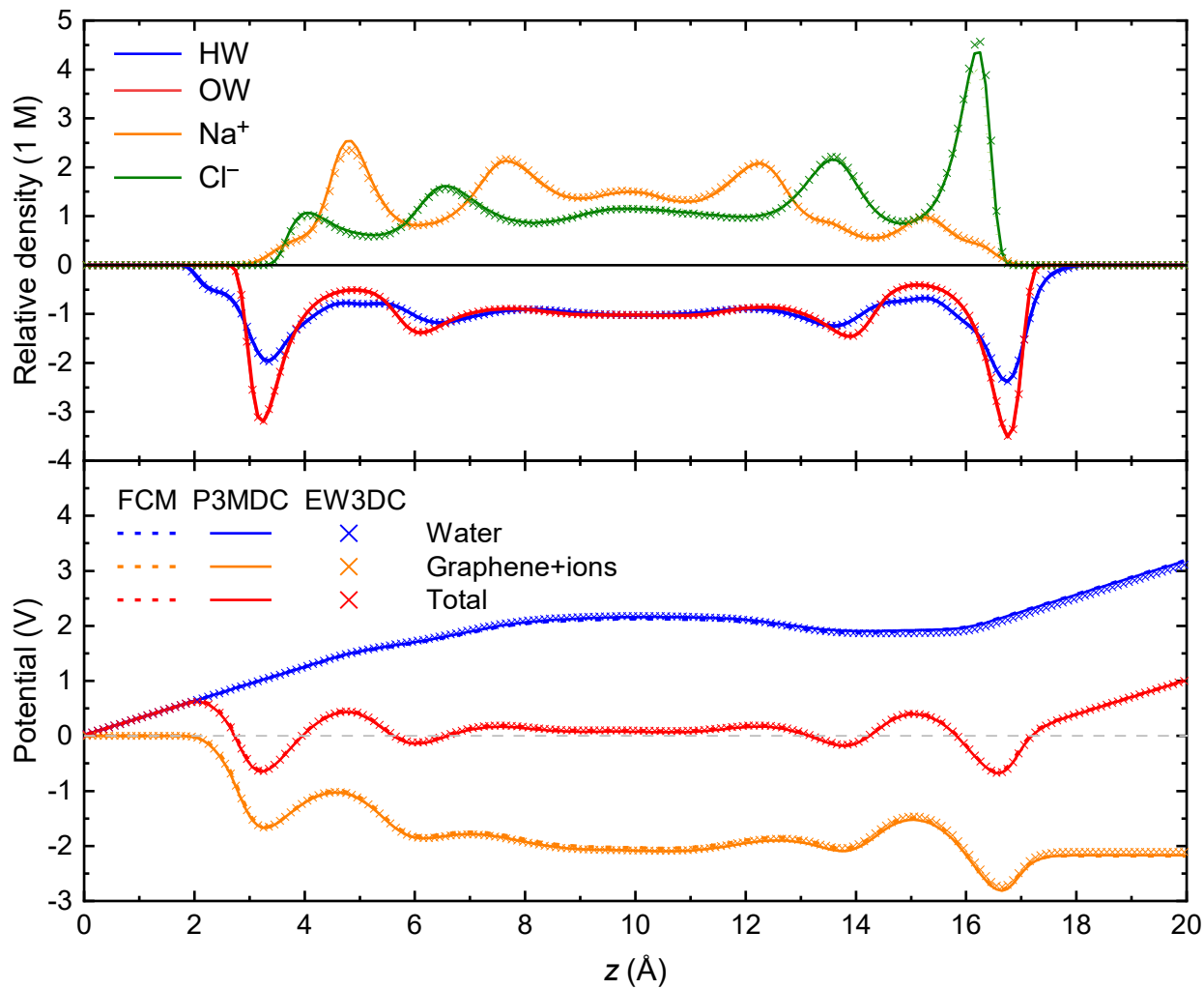

Figure 8: The profiles along *z*-direction in 2-nm graphene simulation cell for (a) EDLs and (b) electrostatic potential in MD simulation. (a) The orange, green, blue, and red curves represent the distribution of Na$^+$, Cl$^-$, hydrogen, and oxygen atoms on water molecules. (b) The net potential (red) is divided into the contributions by water molecules (blue) or the interaction between the graphene charges and the ions (orange). Note that the results by FCM, our CPM work (P3MDC) and Wang's CPM (EW3DC) implementation are shown as dash, solid, and cross, respectively.

Table 1: A summary of SR-CPM parameters for the different 2D lattices

| Lattice | $R_{1nn}$ | Density | $U^{LATT}$ | $\alpha_i^0$ | $\alpha_{ii}^0$ | $\text{erfc}(\alpha_{ii}R_{1nn})$ |
|---|---|---|---|---|---|---|
| **SQ** | 1.000 | 1.000 | 3.90026492000196† | 4.888 | 3.457 | 0.0000010 |
| **HEX** | 1.000 | 1.155 | 4.21342263613691† | 5.281 | 3.734 | 0.0000001 |
| **Kagome** | 1.000 | 0.866 | 3.51118553011409 | 4.401 | 3.112 | 0.0000107 |
| **GRA** | 1.000 | 0.770 | 3.32302166466011 | 4.165 | 2.945 | 0.0000312 |
| **SQ** | 1.619 | 0.382 | 2.40987666959325 | 3.02 | 2.136 | 0.0000010 |
| **HEX** | 1.739 | 0.382 | 2.42270787123462 | 3.036 | 2.147 | 0.0000001 |
| **Kagome** | 1.506 | 0.382 | 2.33125173542534 | 2.922 | 2.066 | 0.0000107 |
| **GRA** | 1.420 | 0.382 | 2.34015610187332 | 2.934 | 2.074 | 0.0000312 |

where $R_{1nn}$ is the distance to the nearest neighbour in the 2D lattice or the bond length; $U^{LATT}$ is the negative electrostatic energy or the Madelung constant for the 2D (Wigner) lattice with $U^{LATT} = \sqrt{2/\pi}\,\alpha_i^0$ and $\alpha_{ii}^0 = \alpha_i^0/\sqrt{2}$. The values in the last column describe the negligible screening effect for the Gaussian charges between the nearest neighbourhood ($R_{1nn}$). Note that the analytic solution of $U^{LATT}$ at $R_{1nn} = 1$ for 2D Square (SQ), Hexagonal (HEX), Kagome, and Graphene (GRA) lattices are $4\zeta(1/2)L_{-4}(1/2)$, $6\zeta(1/2)L_{-3}(1/2)$, $5\zeta(1/2)L_{-3}(1/2)$, and $(3+\sqrt{3})\zeta(1/2)L_{-3}(1/2)$ respectively where $\zeta(x)$ is the Riemann zeta function, and $L_N(x)$ is the Dirichlet *L*-function. The values of square and hexagonal lattices † are taken from Ref. 39.

# References

[1] B. Conway, *Electrochemical supercapacitors: scientific fundamentals and technological applications (POD)* (Kluwer Academic/Plenum: New York, 1999),

[2] J. Vatamanu, O. Borodin, and G. D. Smith, J. Phys. Chem. B **115** (2011) 3073.

[3] C. Merlet, C. Péan, B. Rotenberg, P. A. Madden, P. Simon, and M. Salanne, J. Phys. Chem. Lett. **4** (2013) 264.

[4] Z. Wang, Y. Yang, D. L. Olmsted, M. Asta, and B. B. Laird, J. Chem. Phys. **141** (2014) 184102.

[5] S. Kondrat, N. Georgi, M. V. Fedorov, and A. A. Kornyshev, Phys. Chem. Chem. Phys. **13** (2011) 11359.

[6] C. Merlet, C. Péan, B. Rotenberg, P. A. Madden, B. Daffos, P. L. Taberna, P. Simon, and M. Salanne, Nat. Comm. **4** (2013) 2701.

[7] C. Prehal, C. Koczwara, N. Jäckel, A. Schreiber, M. Burian, H. Amenitsch, M. A. Hartmann, V. Presser, and O. Paris, Nat. Energy **2** (2017) 1.

[8] S. Kondrat, P. Wu, R. Qiao, and A. A. Kornyshev, Nat. Mater. **13** (2014) 387.

[9] A. J. Bard, and L. R. Faulkner, *Electrochemical Methods: Fundamentals and Applications, 2nd Edition* (John Wiley & Sons, Incorporated, 2000),

[10] J. I. Siepmann, and M. Sprik, J. Chem. Phys. **102** (1995) 511.

[11] S. K. Reed, O. J. Lanning, and P. A. Madden, J. Chem. Phys. **126** (2007)

[12] A. V. Raghunathan, and N. R. Aluru, Phys. Rev. E **76** (2007) 011202.

[13] S. Tyagi, A. Arnold, and C. Holm, J. Chem. Phys. **127** (2007) 154723.

[14] M. K. Petersen, R. Kumar, H. S. White, and G. A. Voth, J. Phys. Chem. C **116** (2012) 4903.

[15] M. Girotto, A. P. Dos Santos, and Y. Levin, J. Chem. Phys. **147** (2017) 074109.

[16] S. Stenberg, B. Stenqvist, C. Woodward, and J. Forsman, Phys. Chem. Chem. Phys. **22** (2020) 13659.

[17] R. Allen, J.-P. Hansen, and S. Melchionna, Phys. Chem. Chem. Phys. **3** (2001) 4177.

[18] S. Tyagi, M. Süzen, M. Sega, M. Barbosa, S. S. Kantorovich, and C. Holm, J. Chem. Phys. **132** (2010) 154112.

[19] D. Li, M. B. Muller, S. Gilje, R. B. Kaner, and G. G. Wallace, Nat. Nanotechnol. **3** (2008) 101.

[20] C. Cheng, G. Jiang, C. J. Garvey, Y. Wang, G. P. Simon, J. Z. Liu, and D. Li, Sci. Adv. **2** (2016)

[21] C. Merlet, B. Rotenberg, P. A. Madden, and M. Salanne, Phys. Chem. Chem. Phys. **15** (2013) 15781.

[22] C. Péan, C. Merlet, B. Rotenberg, P. A. Madden, P.-L. Taberna, B. Daffos, M. Salanne, and P. Simon, ACS Nano **8** (2014) 1576.

[23] D. T. Limmer, C. Merlet, M. Salanne, D. Chandler, P. A. Madden, R. Van Roij, and B. Rotenberg, Phys. Rev. Lett. **111** (2013) 106102.

[24] T. R. Gingrich, and M. Wilson, Chem. Phys. Lett. **500** (2010) 178.

[25] J. Vatamanu, D. Bedrov, and O. Borodin, Mol. Simul. **43** (2017) 838.

[26] T. Liang, B. Devine, S. R. Phillpot, and S. B. Sinnott, J. Phys. Chem. A **116** (2012) 7976.

[27] L. Scalfi, D. T. Limmer, A. Coretti, S. Bonella, P. A. Madden, M. Salanne, and B. Rotenberg, Phys. Chem. Chem. Phys. **22** (2020) 10480.

[28] A. Marin-Laflèche, M. Haefele, L. Scalfi, A. Coretti, T. Dufils, G. Jeanmairet, S. K. Reed, S. Alessandra, R. Berthin, and C. Bacon, J. Open Source Softw. Process. **5** (2020) 2373.

[29] A. Coretti, S. Bonella, and G. Ciccotti, J. Chem. Phys. **149** (2018) 191102.

[30] A. Coretti, L. Scalfi, C. Bacon, B. Rotenberg, R. Vuilleumier, G. Ciccotti, M. Salanne, and S. Bonella, J. Chem. Phys. **152** (2020) 194701.

[31] T. Dufils, G. Jeanmairet, B. Rotenberg, M. Sprik, and M. Salanne, Phys. Rev. Lett. **123** (2019) 195501.

[32] T. Dufils, M. Sprik, and M. Salanne, J. Phys. Chem. Lett. **12** (2021) 4357.

[33] L. Pastewka, T. T. Järvi, L. Mayrhofer, and M. Moseler, Phys. Rev. B **83** (2011) 165418.

[34] H. Nakano, and H. Sato, J. Chem. Phys. **151** (2019) 164123.

[35] L. Scalfi, T. Dufils, K. G. Reeves, B. Rotenberg, and M. Salanne, J. Chem. Phys. **153** (2020) 174704.

[36] J. Oshiki, H. Nakano, and H. Sato, J. Chem. Phys. **154** (2021) 144107.

[37] J. N. Israelachvili, *Intermolecular and surface forces* (Academic press, 2015),

[38] E. R. Smith, Proc. Math. Phys. Eng. Sci. **375** (1981) 475.

[39] J. M. Borwein, M. Glasser, R. McPhedran, J. Wan, and I. Zucker, *Lattice sums then and now* (Cambridge University Press, 2013), 150

[40] S. W. de Leeuw, J. W. Perram, and E. R. Smith, Proc. Math. Phys. Eng. Sci. **373** (1980) 27.

[41] J. Lekner, Physica A: Statistical Mechanics and its Applications **157** (1989) 826.

[42] A. Arnold, and C. Holm, Comput. Phys. Commun. **148** (2002) 327.

[43] A. Grzybowski, and A. Bródka, Mol. Phys. **100** (2002) 635.

[44] A. Grzybowski, and A. Bródka, Mol. Phys. **100** (2002) 1017.

[45] D. Shamshirgar, J. Bagge, and A.-K. Tornberg, J. Chem. Phys. **154** (2021) 164109.

[46] V. Rokhlin, J. Comp. Phys. **60** (1985) 187.

[47] M. P. Allen, and D. J. Tildesley, *Computer simulation of liquids* (Oxford university press, 2017),

[48] R. Hockney, and J. Eastwood, *Computer simulation using particles* (Bristol: Hilger, 1988),

[49] T. Darden, D. York, and L. Pedersen, J. Chem. Phys. **98** (1993) 10089.

[50] U. Essmann, L. Perera, M. L. Berkowitz, T. Darden, H. Lee, and L. G. Pedersen, J. Chem. Phys. **103** (1995) 8577.

[51] Y. Shan, J. L. Klepeis, M. P. Eastwood, R. O. Dror, and D. E. Shaw, J. Chem. Phys. **122** (2005) 054101.

[52] F. Hedman, and A. Laaksonen, Chem. Phys. Lett. **425** (2006) 142.

[53] F. Nestler, M. Pippig, and D. Potts, J. Comp. Phys. **285** (2015) 280.

[54] D. Lindbo, and A.-K. Tornberg, J. Comp. Phys. **230** (2011) 8744.

[55] B. R. Brooks, R. E. Bruccoleri, B. D. Olafson, D. J. States, S. a. Swaminathan, and M. Karplus, J. Comp. Chem. **4** (1983) 187.

[56] W. L. Jorgensen, D. S. Maxwell, and J. Tirado-Rives, J. Am. Chem. Soc. **118** (1996) 11225.

[57] D. Van Der Spoel, E. Lindahl, B. Hess, G. Groenhof, A. E. Mark, and H. J. Berendsen, J. Comp. Chem. **26** (2005) 1701.

[58] D. A. Case, T. E. Cheatham III, T. Darden, H. Gohlke, R. Luo, K. M. Merz Jr, A. Onufriev, C. Simmerling, B. Wang, and R. J. Woods, J. Comp. Chem. **26** (2005) 1668.

[59] J. Foresman, and E. Frish, *Exploring chemistry* (Gaussian Inc., Pittsburg, USA, 1996),

[60] P. T. Kiss, M. Sega, and A. Baranyai, J. Chem. Theory Comput. **10** (2014) 5513.

[61] S. Bureekaew, S. Amirjalayer, M. Tafipolsky, C. Spickermann, T. K. Roy, and R. Schmid, Phys. Status Solidi B **250** (2013) 1128.

[62] H. Wei, R. Qi, J. Wang, P. Cieplak, Y. Duan, and R. Luo, J. Chem. Phys. **153** (2020) 114116.

[63] D. Heyes, and A. Brańka, J. Chem. Phys. **138** (2013) 034504.

[64] H. Lee, and W. Cai, Lecture Notes, Stanford University **3** (2009) 1.

[65] T. L. Beck, *Notes on Ewald summation for charges and dipoles,* https://homepages.uc.edu/~becktl/tlb_ewald.pdf (2010)

[66] A. Arnold, and C. Holm, *Efficient Methods to Compute Long-Range Interactions for Soft Matter Systems,* chapter in Advanced Computer Simulation Approaches for Soft Matter Sciences II, edited by C. Holm, and K. Kremer (Springer Berlin Heidelberg, Berlin, Heidelberg, 2005), pp. 59.

[67] J. Kolafa, and J. W. Perram, Mol. Simul. **9** (1992) 351.

[68] M. Deem, J. M. Newsam, and S. Sinha, J. Phys. Chem. **94** (1990) 8356.

[69] J. M. Haile, I. Johnston, A. J. Mallinckrodt, and S. McKay, Comput. Phys. **7** (1993) 625.

[70] A. Y. Toukmaji, and J. A. Board Jr, Comput. Phys. Commun. **95** (1996) 73.

[71] D. Wolf, P. Keblinski, S. Phillpot, and J. Eggebrecht, J. Chem. Phys. **110** (1999) 8254.

[72] T. M. Nymand, and P. Linse, J. Chem. Phys. **112** (2000) 6152.

[73] R. Johnson, and S. Ranganathan, Phys. Rev. E **75** (2007) 056706.

[74] P. J. in't Veld, A. E. Ismail, and G. S. Grest, J. Chem. Phys. **127** (2007) 144711.

[75] O. Osychenko, G. Astrakharchik, and J. Boronat, Mol. Phys. **110** (2012) 227.

[76] T. Prasanna, Philos. Mag. Lett. **92** (2012) 29.

[77] Z. Hu, J. Chem. Theory Comput. **10** (2014) 5254.

[78] B. A. Wells, and A. L. Chaffee, J. Chem. Theory Comput. **11** (2015) 3684.

[79] R. D. Skeel, Mol. Phys. **114** (2016) 3166.

[80] D. Frenkel, and B. Smit, *Understanding molecular simulation: from algorithms to applications* (Elsevier, 2001), Vol. 1,

[81] D. Borwein, J. M. Borwein, and K. F. Taylor, J. Math. Phys. **26** (1985) 2999.

[82] E. Wigner, Phys. Rev. **46** (1934) 1002.

[83] M. Bonitz, P. Ludwig, H. Baumgartner, C. Henning, A. Filinov, D. Block, O. Arp, A. Piel, S. Käding, Y. Ivanov, A. Melzer, H. Fehske, and V. Filinov, Physics of Plasmas **15** (2008) 055704.

[84] O. Emersleben, Phys. Z **24** (1923) 73.

[85] F. Hund, Zeitschrift für Physik **94** (1935) 11.

[86] D. Heyes, J. Chem. Phys. **74** (1981) 1924.

[87] J. Hautman, and M. L. Klein, Mol. Phys. **75** (1992) 379.

[88] B. Stenqvist, New J. Phys. **21** (2019) 063008.

[89] S. Stenberg, and B. Stenqvist, J. Phys. Chem. A **124** (2020) 3943.

[90] S. Boresch, and O. Steinhauser, Berichte der Bunsengesellschaft für physikalische Chemie **101** (1997) 1019.

[91] L. J. Ahrens-Iwers, and R. H. Meißner, J. Chem. Phys. **155** (2021) 104104.

[92] A. Arnold, F. Fahrenberger, C. Holm, O. Lenz, M. Bolten, H. Dachsel, R. Halver, I. Kabadshow, F. Gähler, and F. Heber, Phys. Rev. E **88** (2013) 063308.

[93] M. Deserno, and C. Holm, J. Chem. Phys. **109** (1998) 7694.

[94] J. Lekner, Journal of Electrostatics **69** (2011) 11.

[95] H. Nakano, T. Yamamoto, and S. Kato, J. Chem. Phys. **132** (2010) 044106.

[96] G. Jiang, C. Cheng, D. Li, and J. Z. Liu, Nano Res. **9** (2016) 174.

[97] K. A. O'Hearn, A. Alperen, and H. M. Aktulga, SIAM Journal on Scientific Computing **42** (2020) C1.

[98] J. C. Maxwell, *A Treatise on Electricity and Magnetism* (Cambridge University Press, 2010),

[99] L. Zeng, T. Wu, T. Ye, T. Mo, R. Qiao, and G. Feng,  (2021)

[100] F. Deißenbeck, C. Freysoldt, M. Todorova, J. Neugebauer, and S. Wippermann, Phys. Rev. Lett. **126** (2021) 136803.

[101] D. Borwein, J. Borwein, and R. Shail, J. Math. Anal. Appl. **143** (1989) 126.

[102] J. Choi, J. J. Dongarra, R. Pozo, and D. W. Walker, *ScaLAPACK: A scalable linear algebra library for distributed memory concurrent computers,* in The Fourth Symposium on the Frontiers of Massively Parallel Computation (IEEE Computer Society, 1992), pp. 120.

[103] V. Ballenegger, J. Chem. Phys. **140** (2014) 161102.

[104] S. Yi, C. Pan, and Z. Hu, J. Chem. Phys. **147** (2017) 126101.

[105] A. P. dos Santos, M. Girotto, and Y. Levin, J. Chem. Phys. **144** (2016) 144103.

[106] D. Ongari, P. G. Boyd, O. Kadioglu, A. K. Mace, S. Keskin, and B. Smit, J. Chem. Theory Comput. **15** (2018) 382.

[107] D. R. Bowler, and T. Miyazaki, Rep. Prog. Phys. **75** (2012) 036503.

[108] A. Serva, L. Scalfi, B. Rotenberg, and M. Salanne, arXiv preprint arXiv:2106.07232 (2021)

[109] A. Serva, L. Scalfi, B. Rotenberg, and M. Salanne, J. Chem. Phys. **155** (2021) 044703.

[110] Y. Matsumi, H. Nakano, and H. Sato, Chem. Phys. Lett. **681** (2017) 80.

[111] I.-C. Yeh, and M. L. Berkowitz, J. Chem. Phys. **111** (1999) 3155.

[112] W. L. Jorgensen, J. Chandrasekhar, J. D. Madura, R. W. Impey, and M. L. Klein, J. Chem. Phys. **79** (1983) 926.

[113] J.-P. Ryckaert, G. Ciccotti, and H. J. C. Berendsen, J. Comp. Phys. **23** (1977) 327.

[114] P. Li, L. F. Song, and K. M. Merz, J. Chem. Theory Comput. **11** (2015) 1645.

[115] W. G. Hoover, Phys. Rev. A **31** (1985) 1695.

[116] F. Iori, and S. Corni, J. Comp. Chem. **29** (2008) 1656.

# Supplementary Information

## Electrostatic potential by Gaussian charge

As mentioned in the main text, the interaction between a point charge and a Gaussian charge, or the electrostatic potential of a unit Gaussian charge at the distance of **r** is given as

$$\varphi(\mathbf{r}; \alpha_i) = \int \frac{G(\mathbf{r}')}{|\mathbf{r} - \mathbf{r}'|} d\mathbf{r}' = \frac{\text{erf}(\alpha_i |\mathbf{r}|)}{|\mathbf{r}|} \tag{S1}$$

where $\alpha_i$ is the inverse distribution width of the Gaussian charge and the centre of Gaussian charge is located on the origin point. The above equation can be obtained by solving the Poisson's equation as $\nabla^2 \varphi(\mathbf{r}) = -4\pi G(\mathbf{r})$ in the Gaussian units. Attributed to the spherical symmetric distribution of $G(\mathbf{r})$, we take it in the spherical coordinate as

$$\frac{1}{r}\frac{\partial^2}{\partial r^2}[r\varphi(r; \alpha_i)] = -4\pi G(r)$$
$$\frac{\partial^2}{\partial r^2}[r\varphi(r; \alpha_i)] = -4\pi r G(r)$$
$$\frac{\partial}{\partial r}[r\varphi(r; \alpha_i)] = 4\pi \int_r^\infty r' G(r') dr' = \frac{2\pi}{\alpha_i^2} G(r) \tag{S2}$$
$$r\varphi(r; \alpha_i) = \frac{2\pi}{\alpha^2}\int_0^r G(r') dr' = \frac{2\pi}{\alpha_i^2}\left(\frac{\alpha_i^2}{\pi}\right)^{3/2}\frac{\sqrt{\pi}}{2\alpha_i}\text{erf}(\alpha_i r)$$
$$\varphi(r; \alpha_i) = \frac{\text{erf}(\alpha_i r)}{r}$$

## Fourier transformation of the potential by a Gaussian charge

Here, we take the Fourier transform of the electrostatic potential of a Gaussian charge in Eq. (S1). By using the Poisson's equation in the reciprocal space $k^2 \hat{\varphi}(\mathbf{r}) = 4\pi \hat{G}(\mathbf{r})$, it yields,

$$\begin{aligned}
\hat{\varphi}(\mathbf{k}, \alpha_i) &= \frac{4\pi}{k^2}\hat{G}(\mathbf{k}; \alpha_i) \\
&= \frac{4\pi}{k^2}\left(\frac{\alpha_i^2}{\pi}\right)^{3/2} \int e^{-\alpha_i^2 x^2 - ik_x x} dx \int e^{-\alpha_i^2 y^2 - ik_y y} dy \int e^{-\alpha_i^2 z^2 - ik_z z} dz \\
&= \frac{4\pi}{k^2}\exp\left(-\frac{k_x^2}{4\alpha_i^2}\right)\exp\left(-\frac{k_y^2}{4\alpha_i^2}\right)\exp\left(-\frac{k_z^2}{4\alpha_i^2}\right) \\
&= \frac{4\pi}{k^2}\exp\left(-\frac{k^2}{4\alpha_i^2}\right)
\end{aligned} \quad (S3)$$

where $k \equiv |\mathbf{k}| = \sqrt{k_x^2 + k_y^2 + k_z^2}$ and the 2$^{\text{nd}}$ to 3$^{\text{rd}}$ lines is obtained through the Fourier transform of a Gaussian distribution as $\int \exp(-\alpha_i^2 x^2 - ikx)dx = \frac{\sqrt{\pi}}{\alpha_i}\exp(k^2/4\alpha_i^2)$. Since the point charge can be regarded as the dimensionless Gaussian charge with $\alpha_i \to \infty$, the Coulomb potential for the point charge is obtained as $\hat{\varphi}(\mathbf{k}) = 4\pi/k^2$.

### Potential between two Gaussian charges

The potential formula, or the Green's function between two unit Gaussian charges, could be solved by taking a Coulomb integral, namely, through the integral of the charge distribution of one Gaussian charge $G(\mathbf{r}; \alpha_i)$ with the electrostatic potential of the other Gaussian charge $\varphi(\mathbf{r}; \alpha_j)$,

$$\varphi(\mathbf{r}_{ij}, \alpha_{ij}) = \int G(\mathbf{r} - \mathbf{r}_i; \alpha_i)\varphi(\mathbf{r} - \mathbf{r}_j; \alpha_j)d\mathbf{r} \quad (S4)$$

Eq. (S4) is a convolution integral in 3D space. We handle it easily in the reciprocal space as,

$$\begin{aligned}
\hat{\varphi}(\mathbf{k}, \alpha_{ij}) &= \hat{G}(\mathbf{k}; \alpha_i)\hat{\varphi}(\mathbf{k}; \alpha_j) \\
&= \frac{k^2}{4\pi}\hat{\varphi}(\mathbf{k}; \alpha_i)\hat{\varphi}(\mathbf{k}; \alpha_j) \\
&= \frac{4\pi}{k^2}\exp\left(-\frac{k^2}{4\alpha_i^2} - \frac{k^2}{4\alpha_j^2}\right) \\
&= \frac{4\pi}{k^2}\exp\left(-\frac{k^2}{4\alpha_{ij}^2}\right)
\end{aligned} \quad (S5)$$

Then, the inverse Fourier transform gives a screen Coulomb charge in the real space as $\hat{\varphi}(\mathbf{k}_{ij}, \alpha_{ij}) = \frac{\text{erf}(\alpha_{ij}r)}{r}$. It has the similar formula as Eq. (S1) for the potential of a Gaussian charge.

## Fourier transformation of Ewald potential

We give an example to take the Fourier transform of the periodic Ewald potential, using the Poisson summation.

$$\begin{aligned}
\hat{\psi}(\mathbf{k}) &= \sum_{\mathbf{n}} \int \frac{e^{i\mathbf{k}\cdot\mathbf{r}}}{|\mathbf{r} - \mathbf{R_n}|} d\mathbf{r} = \sum_{\mathbf{n}} e^{i\mathbf{k}\cdot\mathbf{R_n}} \int \frac{e^{i\mathbf{k}\cdot\mathbf{r}'}}{|\mathbf{r}'|} d\mathbf{r}' \\
&= \frac{4\pi}{k^2} \sum_{\mathbf{n}} \exp\left(i2\pi\mathbf{k} \cdot \frac{\mathbf{R_n}}{2\pi}\right) \\
&= \frac{8\pi^3}{V} \frac{4\pi}{k^2} \sum_{\mathbf{K}} \delta(\mathbf{k} - \mathbf{K})
\end{aligned} \quad (S6)$$

The last line of Eq. (S6) is obtained through the Poisson summation formula, giving the *Dirac comb* from $\sum_n e^{i2\pi k \frac{n}{T}} = \sum_n T\, \delta(k - nT)$ where $T = \frac{2\pi}{L}$ and $\mathbf{K} = 2\pi \mathbf{L}^{-1} \circ \mathbf{n}$.

## Madelung constant in Ewald interaction matrix

The particular feature of the Ewald interaction matrix is the non-zero self-interaction in the diagonal, $e_{ii} = \psi(\mathbf{0})$, which is attributed to the interaction of $q_i$ with all its images as well as the neutralising background. As shown in our pedagogical example of three point-charges in a cubic periodic cell (Fig. 1), the diagonal elements versus the cell dimension $L$ is found to be a constant $\xi = \psi(\mathbf{0})/L = -2.837297479$, which is sometimes named as the Madelung constant of a Simple Cubic (Wigner) crystal, $U^{SC}$.[1]

The non-zero self-interaction energy indicates the Ewald potential could be divided into two parts as the zero-separation or baseline potential $\psi(\mathbf{0})$, and the increment potential by the position displacement $\Delta\psi(\mathbf{r})$. Once $r \ll L$, the Coulomb's law is recovered (Fig. 1)

$$\lim_{r \to 0} \Delta \psi(\mathbf{r}) = \lim_{r \to 0} \psi(\mathbf{r}) - \psi(\mathbf{0})$$
$$= \lim_{r \to 0} \psi^0(\mathbf{r}) + \frac{1}{|\mathbf{r}|} - \frac{\text{erf}(\eta |\mathbf{r}|)}{|\mathbf{r}|} - \psi^0(\mathbf{0}) + \frac{2\eta}{\sqrt{\pi}} \quad \text{(S7)}$$
$$\approx \frac{1}{|\mathbf{r}|}$$

Similarly, we divide **E** into two parts, the symmetry *hollow* matrix (zero diagonal entries) for the inter-particle interaction plus an *all-ones* matrix multiplied with the Madelung constant $\xi$.[2] The former illustrates the potential increment by the separation of two charge (**r**). Oppositely, the *all-ones* matrix only contributes in the non-neutral simulation at $Q \neq 0$. For the Gaussian charges, once $L \gg \alpha_i^{-1}$, the following approximation holds as $\psi(\mathbf{0}; \alpha_{ii}) \approx \psi(\mathbf{0}; \infty) + \varphi(\mathbf{0}; \alpha_{ii})$. $\varphi(\mathbf{0}; \alpha_{ii}) = \sqrt{1/\pi}\alpha_{ii}$ is the self-energy of the Gaussian charge.[3] and $\psi(\mathbf{0}; \infty)$ is the self-interaction energy of a point charge.

It is worth noting that $\xi$ still plays an essential role in plenty of fields involving the periodic pattern, *e.g.*, the accounting of the ionic hydration energy in a periodic cell[4], the size-dependent effect in MD simulation[5], or even in nuclear physics research[6]. Moreover, if dealing with the other shapes of the simulation cells or the periodic patterns, the other Madelung constants might be involved, *e.g.*, $U^{\text{FCC}} = -4.58536207398061$ and $U^{\text{BCC}} = -3.63923345028097$ for the Face-Centred Cubic (FCC) or Body-Centred Cubic (BCC) cells, respectively.[1] Note that such terminology is slightly different from the original definition of the ionic crystal, *e.g.*, rock salt with the alternative arrangement of both cations and anions.

### Neutralising vector d in the iterative approaches

As shown in Fig. 5, electroneutrality plays a crucial role in accurately describing the polarisability of the conducting metal. In this work, we enforce the constraint through a general post-treatment step. It involves a neutralising vector **d** from **E**, as

$$d_i = \frac{\sum_j e_{ij}^{-\hat{1}}}{\sum_{i,j} e_{ij}^{-\hat{1}}} \quad \text{(S8)}$$

Where $e_{ij}^{-\hat{1}}$ is the entry of $\mathbf{E}^{-1}$. In the iterative approach, we circumvent the calculation of $\mathbf{E}^{-1}$ by solving the following equation iteratively.

$$\mathbf{E}\mathbf{d}' = \mathbf{1} \qquad (S9)$$

where $\mathbf{1}$ is an all-ones vector as $\mathbf{1} = (1,1,\ldots,1)^T$. Thus, each entry in $\mathbf{d}'$ is the sum of one column of $\mathbf{E}^{-1}$ as $d'_i = \sum_j e_{ij}^{-1}$. The neutralising vector $\mathbf{d}$ is obtained by $d_i = d'_i / \sum_j d'_j$.

## Charge matrix of $\widetilde{\mathbf{Q}}$ and $\widetilde{\mathbf{Q}}^{-1}$ in MI method

As mentioned in **Sec. IIIB.1**, the invertible matrix $\widetilde{\mathbf{Q}}$ is constructed in advance with the same diagonal and off-diagonal elements, which are constituted by an identity matrix $\mathbf{I}$ and an $N \times N$ all-ones matrix $\mathbf{1}_{N \times N}$

$$\widetilde{\mathbf{Q}} = a\mathbf{I} + b\mathbf{1}_{N \times N} \qquad (S10)$$

where $a$ and $b$ are the scalar coefficients. The inverted matrix $\widetilde{\mathbf{Q}}^{-1}$ should have similar components as

$$\widetilde{\mathbf{Q}}^{-1} = c\mathbf{I} + d\mathbf{1}_{N \times N} \qquad (S11)$$

The multiplication with each other equals the identity matrix

$$\begin{aligned}\mathbf{I} &= \widetilde{\mathbf{Q}} \cdot \widetilde{\mathbf{Q}}^{-1} \\ &= ac\mathbf{I} + (bc + ad)\mathbf{1}_{N \times N} + bd(\mathbf{1}_{N \times N} \cdot \mathbf{1}_{N \times N}) \\ &= ac\mathbf{I} + (bc + ad + Nbd)\mathbf{1}_{N \times N}\end{aligned} \qquad (S12)$$

where $\mathbf{1}_{N \times N}^k = N^{k-1}\mathbf{1}_{N \times N}$ and $\exp(b\mathbf{1}_{N \times N}) = \sum_k \frac{b^k}{k!}\mathbf{1}_{N \times N}^k = \frac{1}{N}\exp(bN) \cdot \mathbf{1}_{N \times N}$. To hold the equality, it should be $ac = 1$ and $bc + ad + Nbd = 0$. In practice, we set $b \equiv -1$ and $a \equiv N + 1$ that yields $c = d = 1/(N+1)$. Namely, the diagonal elements is double of the off-diagonal ones in $\widetilde{\mathbf{Q}}^{-1}$. The example in Fig. 1 are double those of 0.5 and 0.25 for the diagonal and off-diagonal entries in $\widetilde{\mathbf{Q}}^{-1}$, respectively. Furthermore, to construct an invertible matrix, the determinant of $\widetilde{\mathbf{Q}}$ must be nonzero,

$$\begin{aligned}
\det(\tilde{\mathbf{Q}}) &= \det(a\mathbf{I} + b\mathbf{1}_{N\times N}) = a^N \det(\mathbf{I} + p\mathbf{1}_{N\times N}) \\
&= a^N \exp\big[\mathrm{tr}\big(\log(\mathbf{I} + p\mathbf{1}_{N\times N})\big)\big] \\
&= a^N \exp\left[\mathrm{tr}\left(\sum_{k=1}^{\infty}(-1)^{k+1}\frac{p^k \mathbf{1}_{N\times N}^k}{k}\right)\right] \\
&= a^N \exp\left[\mathrm{tr}\left(\frac{1}{N}\sum_{k=1}^{\infty}(-1)^{k+1}\frac{(pN)^k}{k}\cdot \mathbf{1}_{N\times N}\right)\right] \\
&= a^N \exp\left[\frac{\log(pN+1)}{N}\mathrm{tr}(\mathbf{1}_{N\times N})\right] \\
&= a^N(pN+1) = a^{N-1}(bN+a)
\end{aligned} \tag{S13}$$

where $p \equiv b/a$. If $b$ is set as $b = -1$, $a$ must be employed as $a \neq N$ or $a \neq 0$.

## Galvanostatic charge-discharge (GCD) simulation

The galvanostatic charge-discharge (GCD) process plays a fundamental role in the various electrochemical applications involving the battery, supercapacitor, and even water desalination. During the charge-discharge process, the total amount of the charge on each electrode is controlled but allows to redistribute on the individual electrode/conductor to minimise the electrostatic energy (like Eq. 24). Recently, Zeng et al.[7] developed a modelling approach to simulate the GCD process under the constant potential constraint (GCD-CPM). Herein, we propose a similar method to simulate the GCD process using our two-step method developed in the main text of **Sec. IIIA.**. The total electrostatic energy under both the total charge constraint (*i.e.* GCD constraint) and isopotential constraint for a system of $M$ electrodes is given as,

$$\begin{aligned}
U\{\mathbf{q}^{\mathrm{ele}}, \mathbf{q}^{\mathrm{sol}}\} = &\frac{1}{2}\sum_{I,J} q_I q_J e_{IJ} + \sum_{i,I} q_i q_I e_{iI}(\alpha_i) + \frac{1}{2}\sum_{i,j} q_i q_j e_{ij}(\alpha_{ij}) \\
&- \sum_i \chi_i q_i - \sum_a^M \Delta\mu_a\left(\sum_{a_i} q_{a_i} - Q_a^0\right)
\end{aligned} \tag{S14}$$

where the capital and lowercase alphabet subscripts of $i, j, k, \ldots$ or $I, J, K, \ldots$, indicate the $i$th electrode atom or the $I$th electrolyte atom, respectively. While, the subscripts of $a, b, c, \ldots$ refer to the $a$th electrode in the simulation. Then the complex subscript, like $a_i$ refers to $i$th atom belongs to $a$the electrode. Thus, the summation of $a_i$ in the last term suggests the accounting of the total charge on the $a$th electrode. The $Q_a^0$ is the given total charge constraint and $\Delta\mu_a$ is the corresponding Lagrange multiplier for $a$th electrode. $\chi_i$ in 4$^{th}$ term is the external potential constraint, which is given to keep consistent with Eq. 24. Attributed to the undetermined $\Delta\mu_a$, the influence of $\chi_i$ is eliminated. Taking the derivative with respect to the electrode charge $q_i$, the equilibrium charge distribution is given as

$$\sum_j q_j e_{ij} = \chi_i - \sum_I q_I e_{iI} + \Delta\mu_i \qquad (S15)$$
$$= b_i + \Delta\mu_i$$

where $\Delta\mu_i \equiv \Delta\mu_a, \forall i \in a$ is the electrochemical shift for $i$th atom on the $a$th electrode. Similar to the electroneutrality constraint (Eq. 31), we solved a system of linear equations as $\mathbf{E}\mathbf{q}^{\text{ele}} = \mathbf{b} + \Delta\mathbf{u}$, which could be divided into the two parts of $\mathbf{q}^b$ and $\mathbf{q}^{\Delta\mu}$ as

$$\mathbf{q}^{\text{ele}} = \mathbf{E}^{-1}\mathbf{q}^b + \mathbf{E}^{-1}\Delta\mathbf{u} \qquad (S16)$$
$$= \mathbf{q}^b + \mathbf{q}^{\Delta\mu}$$

or specifically,

$$q_i = \sum_j e_{ij}^{-\hat{1}} b_j + \sum_j e_{ij}^{-\hat{1}} \Delta\mu_j$$
$$= \sum_j e_{ij}^{-\hat{1}} b_j + \sum_b \Delta\mu_b \sum_{b_j} e_{ib_j}^{-\hat{1}} \qquad (S17)$$
$$= q_i^b + q_i^{\Delta\mu}$$

where $\Delta\mathbf{u} = (\Delta\mu_1, \Delta\mu_2, \dots, \Delta\mu_N)^T$ is the electrochemical potential shift vector for $N$ electrode atoms, and $\mathbf{q}^b \equiv (q_1^b, q_2^b, \dots, q_N^b)^T$ and $\mathbf{q}^{\Delta\mu} \equiv (q_1^{\Delta\mu}, q_2^{\Delta\mu}, \dots, q_N^{\Delta\mu})^T$ is the corresponding charge vectors by $\mathbf{b}$ and $\Delta\mathbf{u}$, respectively. Under the total charge constraint, the sum of $q_i^{\Delta\mu}$ for each electrode is given as $\Delta Q_a = Q_a^0 - \sum_{a_i} q_{a_i}^b = \sum_{a_i} q_{a_i}^{\Delta\mu}$. From Eq. (S17), the relation between $\Delta Q_a$ and $\Delta\mu_b$ is given as,

$$\Delta Q_a = \sum_b \Delta\mu_b \sum_{a_i, b_j} e_{a_i b_j}^{-\hat{1}}$$
$$= \sum_b C_{ab} \Delta\mu_b \qquad (S18)$$

where $C_{ab} = \sum_{a_i,b_j} e^{-\hat{1}}_{a_i b_j}$ is the capacitance coefficient between $a$th and $b$th electrodes[8], which equals the sum of the entries of $e^{-1}_{ij}$ in **E** between all atoms of $i$ and $j$ belongs to the $a$th and the $b$th electrodes, respectively. Thus, Eq. (S18) describes the overall charging behaviour of a system of $M$ electrodes, yielding the matrix equation of,

$$\Delta \mathbf{q}_E = \mathbf{C}_E \Delta \mathbf{u}_E \quad (S19)$$

where $\mathbf{C}_E$ is $M \times M$ capacitance coefficient matrix between all $M$ electrodes. The entries of $C_{ab}$ is summed (coarse-grained) from **E**. $\Delta \mathbf{q}_E = (\Delta Q_1, \Delta Q_2, \ldots, \Delta Q_M)^T$ is the electrode charge vector and $\Delta \mathbf{u}_E = (\Delta \mu_1, \Delta \mu_2, \ldots, \Delta \mu_M)^T$ is the electrochemical potential shift vector. The subscript $E$ indicates the matrix or the vector is coarse-grained from the atomic counterpart. Once $\Delta \mathbf{q}_E$ is determined from $\mathbf{q}^b = \mathbf{E}^{-1}\mathbf{b}$, we could easily determine the electrochemical potential shift on each electrode by

$$\Delta \mathbf{u}_E = \mathbf{C}_E^{-1} \Delta \mathbf{q}_E \quad (S20)$$

where $\mathbf{C}_E^{-1}$ is the inverse of $\mathbf{C}_E$. Then, $\mathbf{q}^{\Delta\mu}$ would be easily derived from $\mathbf{E}^{-1}$ and $\Delta \mathbf{u}_E$

For the conventional electrochemical application with two electrodes of $a$ and $b$, $\mathbf{C}_E$ is a $2 \times 2$ matrix with three different capacitance coefficients $C_{aa}$, $C_{bb}$, and $C_{ab}$. $C_{aa}$ is the capacitance of electrode $a$ at the given potential ($V_a = V$) but the other electrode $b$ is ground ($V_b = 0$), which is defined as $C_{aa} \equiv Q_a/V$. Whereas, $C_{ab}$ is the (negative) capacitance of electrode $a$ at $V_a = 0$ and $V_b = V$ defined as $C_{ab} \equiv Q_a/V < 0$

The subtle difference between $C_{aa}$ and $C_{ab}$ is attributed to the third invisible "grounded" electrode, *infinity*. Namely, the electric field lines emerging from the electrode $a$ would terminate at both electrode $b$ and *infinity*. Unlike the infinite parallel plate capacitor, the electrodes with a finite dimension, such as the two spherical electrodes, would suffer the influence of the third electrode at infinity ($r \to \infty$). In this case at $V_a = 0$ and $V_b = V$, the amount of the induced charges on both electrodes are not equal ($Q_a \neq -Q_b$), which is given by $Q_b = \frac{C_{ab}}{C_{aa}} Q_a$.[9] Under the PBCs, the neutralising background plays the role of the third electrode.

Moreover, the joint capacitance of two electrodes with the opposite charges $Q$ is given as[9]

$$C(Q,-Q) = \frac{Q}{V_a - V_b} = \frac{C_{aa}C_{bb} - C_{ab}^2}{C_{aa} + 2C_{ab} + C_{bb}} = \frac{|\mathbf{C}_E|}{C^{\text{ele}}} \tag{S21}$$

where $|\mathbf{C}_E|$ is the determinant of the capacitance coefficients matrix and $C^{\text{ele}} \equiv \sum_{ij} e_{ij}^{-\hat{1}}$ is the total capacitance of the system of all electrodes at the same potential. Note that the formula of $C(Q,-Q)$ is same as the capacitance in vacuum of $C_{\text{diff}}^{\text{empty}}$ proposed in Ref. 10.

## Iterative approach for GCD-CPM

In the previous section, a two-step method was developed to simulate the GCD process. It is similar to the recently proposed method by Zeng et al.[7], using a matrix approach with a $(N+2) \times (N+2)$ matrix. However, the abundant matrix-vector multiplications in the matrix approach would limit the application to the extensive system ($N > 10^4$). As a result, we developed a consistent two-step method to apply GCD-CPM simulation in the iterative techniques without the calculation of $\mathbf{E}^{-1}$ or the matrix-vector multiplication. The capacitance coefficient matrix $\mathbf{C}_E$ is first calculated via iterative techniques. Herein we summarise the calculation method in the following three steps,

1. The $h$th electrode is selected with a given atomic potential vector $\mathbf{b}^h$,

$$b_j^h = \begin{cases} 1 & j \in h \\ 0 & j \notin h \end{cases} \tag{S22}$$

2) The system of linear equations of $\mathbf{E}\mathbf{a}^h = \mathbf{b}^h$ for $N$ electrode atom is solved using iterative methods, giving $\mathbf{a}^h = \mathbf{E}^{-1}\mathbf{b}^h$,

$$a_i^h = \sum_j e_{ij}^{-\hat{1}} b_j^h = \sum_{b_j} e_{ib_j}^{-\hat{1}} \tag{S23}$$

3) Repeat steps (1)-(2) by $M$ times for all electrodes, A $N \times M$ matrix $\mathbf{A}$ is assembled as $\mathbf{A} = (\mathbf{a}^1, \mathbf{a}^2, \ldots, \mathbf{a}^M)$. Thus, $\mathbf{C}_E$ could be derived as

$$C_{ab} = \sum_{a_i} a_{a_i}^b = \sum_{a_i} \sum_{b_j} e_{a_i b_j}^{-1} \tag{S24}$$

Note that once $\mathbf{A}$ is obtained from the iterative approach, $\mathbf{q}^{\Delta\mu}$ could be determined from $\mathbf{q}^{\Delta\mu} = \mathbf{A}\Delta\mathbf{u}$.

## Potential by a Gaussian charge using the P3M solver

The FFT techniques, *e.g.*, P3M, PME, and SPME have significantly accelerated the Ewald summation in the reciprocal space with a slight scarification of the accuracy. Herein, we studied the accuracy of electrostatic interactions between the point and Gaussian charges, using P3M with different accuracy settings. The width of the Gaussian charge was set to be $\alpha_i = 1$ Å$^{-1}$ and the real space cut is $r_{cut} = 10$ Å. Figure S1a&c show the change of the forces (a) and energy (b) for a pair of charges over a range from 1 to 16 Å in a 100 Å cubic box. All the pairs of interactions, including point-to-point (open), point-to-Gaussian (half), and Gaussian-to-Gaussian (solid), are consistent with the analytic result. The electrostatic force between two Gaussian charges is given by

$$\mathbf{F}^G(\mathbf{r}; \alpha_{ij}) = -\nabla \varphi^G(\mathbf{r}; \alpha_{ij}) = -\nabla \frac{\text{erf}(\alpha_{ij} r)}{r}$$
$$= \left( \frac{\text{erf}(\alpha_{ij} r)}{r^2} - \frac{2\alpha_i}{\sqrt{\pi}} \frac{e^{-\alpha_i^2 r^2}}{r} \right) \frac{\mathbf{r}}{r} \quad \text{(S25)}$$

Figure S1b&d illustrate the relative error for force (a) and energy (b) respectively. Although the relative error increased with the separation, the absolute magnitudes decreased. Thus, the larger relative error at the long-range contributes negligibly. It is worth noting that the width of the Gaussian charge is much smaller than the separation distance. Thus, only at a very close range (< 2 Å), the results of the Gaussian-to-interaction become distinctively different from that of the point-to-point one. Such $\alpha_i = 1$ Å$^{-1}$ was taken by considering the optimal $\alpha_i^0$ proposed in Table 1. The similar tendency of the relative error between different kinds of interactions suggests that the magnitude of the error is mainly attributed to the intrinsic trade-off by the P3M method. We also discovered a moderate jump of the error at $r = 10$ Å, which is attributed to the real space truncation (dashed line).

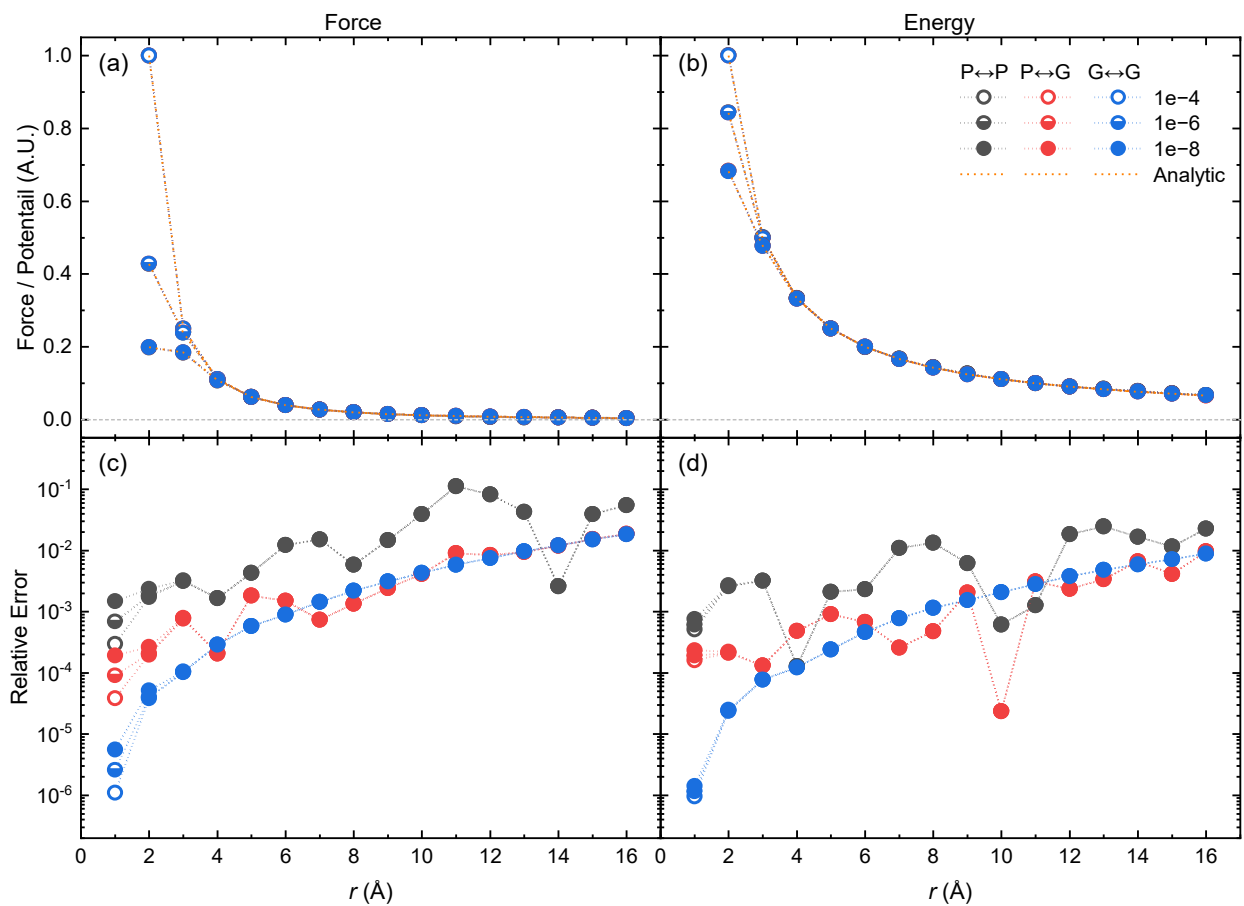

Figure S1: The pairwise point-to-point (open), point-to-Gaussian (half), and Gaussian-to-Gaussian (solid) electrostatic interaction of the force (left) and energy (right). The bottom two figures illustrate the corresponding relative errors. The calculation between two charges was taken in a cubic 100 Å simulation cell, using the P3M method. The distribution width of the Gaussian charge is set to be $\alpha_i = 1$ Å$^{-1}$. The P3M screening factor $\eta$ and the grid number in one dimension for different accuracy (*i.e.*, $10^{-4}, 10^{-6}, 10^{-8}$) are 0.127782, 0.241249, 0.323299 Å$^{-1}$ and 24, 80, 225 respectively. The orange dashed lines are the analytical results.

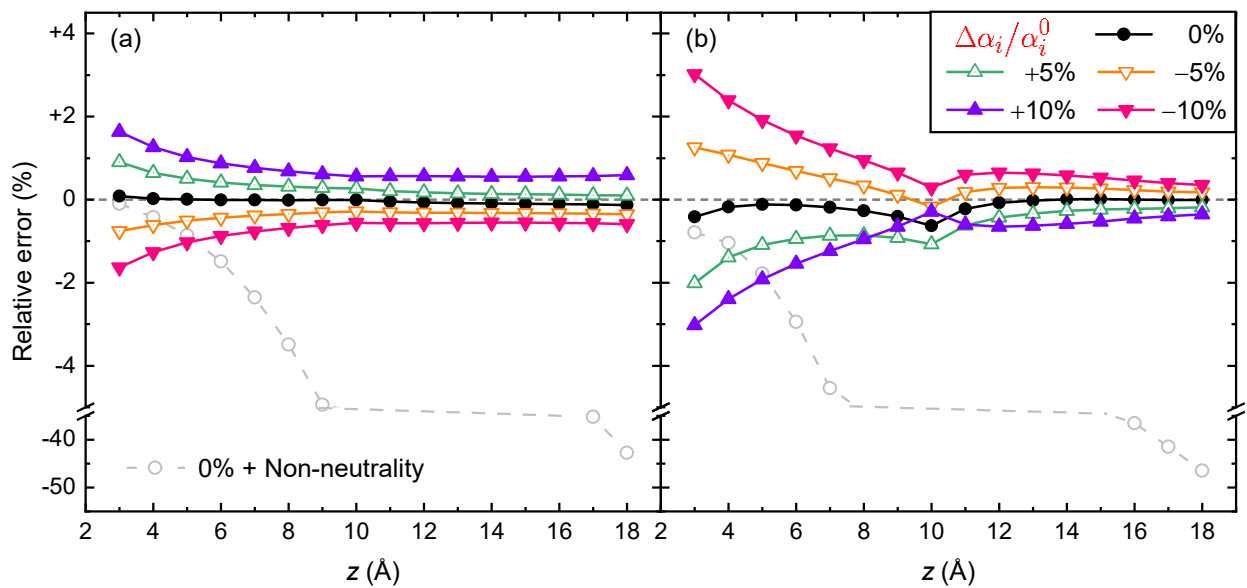

Figure S2: The mean and spread of the relative errors of energy (a) and force in $z$-direction (b) versus the distance $\Delta z$ of the point charges to the 2D square sheet. The lateral sizes of the sheets are 103.43 × 103.43 Å and grounded by the CPM method. The percentages in the right top legend describe the magnitude of the deviation from the optimal $\alpha_i^0 = 3.024$ Å$^{-1}$ at $a = 1.616$ Å. The grey dashed line is the result without the electroneutrality constraint.

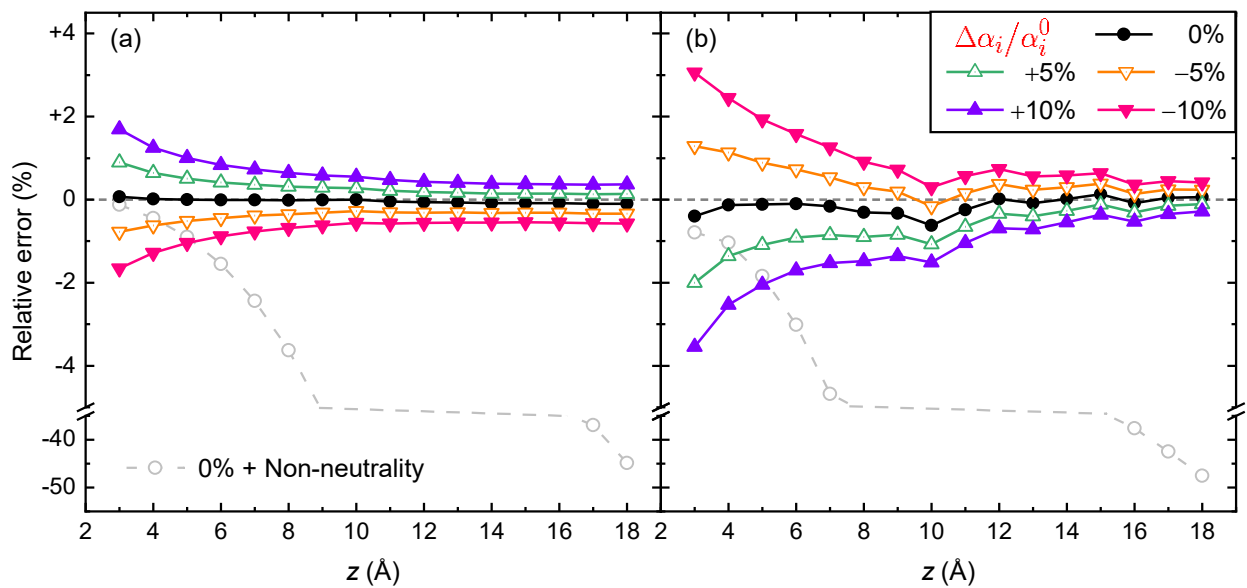

Figure S3: The mean and spread of the relative errors of energy (a) and force in *z*-direction (b) versus the distance $\Delta z$ of the point charges to the 2D hexagonal sheet. The lateral sizes of the sheets are 100.728 × 102.273 Å and grounded by the CPM method. The percentages in the right top legend describe the magnitude of the deviation from the optimal $\alpha_i^0 = 3.041$ Å$^{-1}$ at $a = 1.736$ Å. The grey dashed line is the result without the electroneutrality constraint.

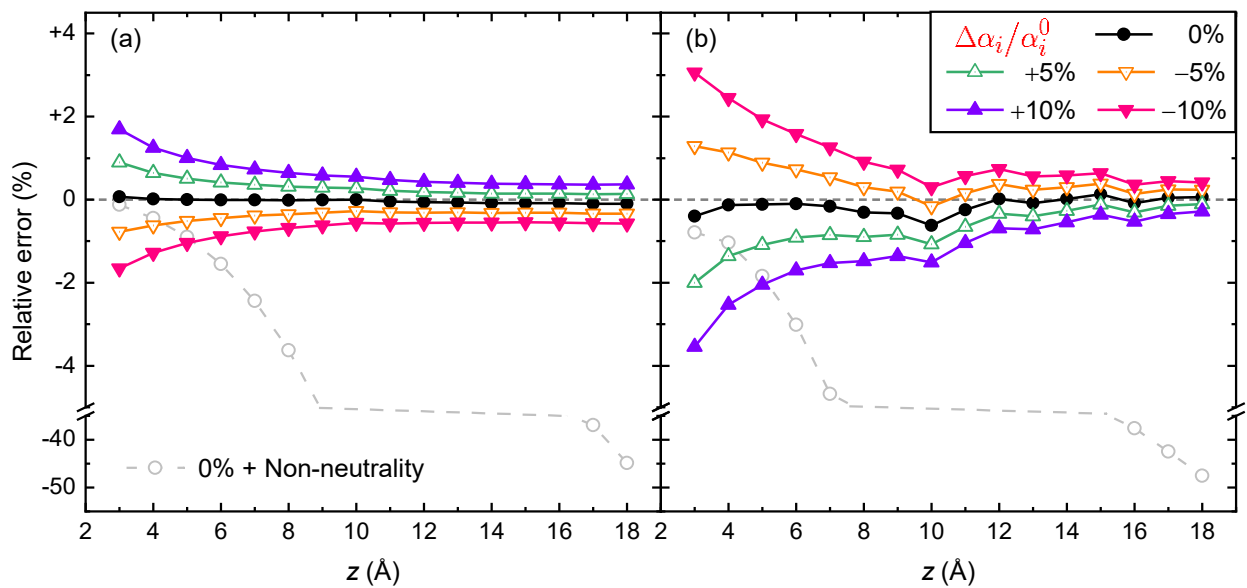

Figure S4: The mean and spread of the relative errors of energy (a) and force in $z$-direction (b) versus the distance $\Delta z$ of the point charges to the 2D Kagome sheet. The lateral sizes of the sheets are 108.442 × 104.348 Å and grounded by the CPM method. The percentages in the right top legend describe the magnitude of the deviation from the optimal $\alpha_i^0 = 2.922$ Å$^{-1}$ at $a = 1.506$ Å. The grey dashed line is the result without the electroneutrality constraint.

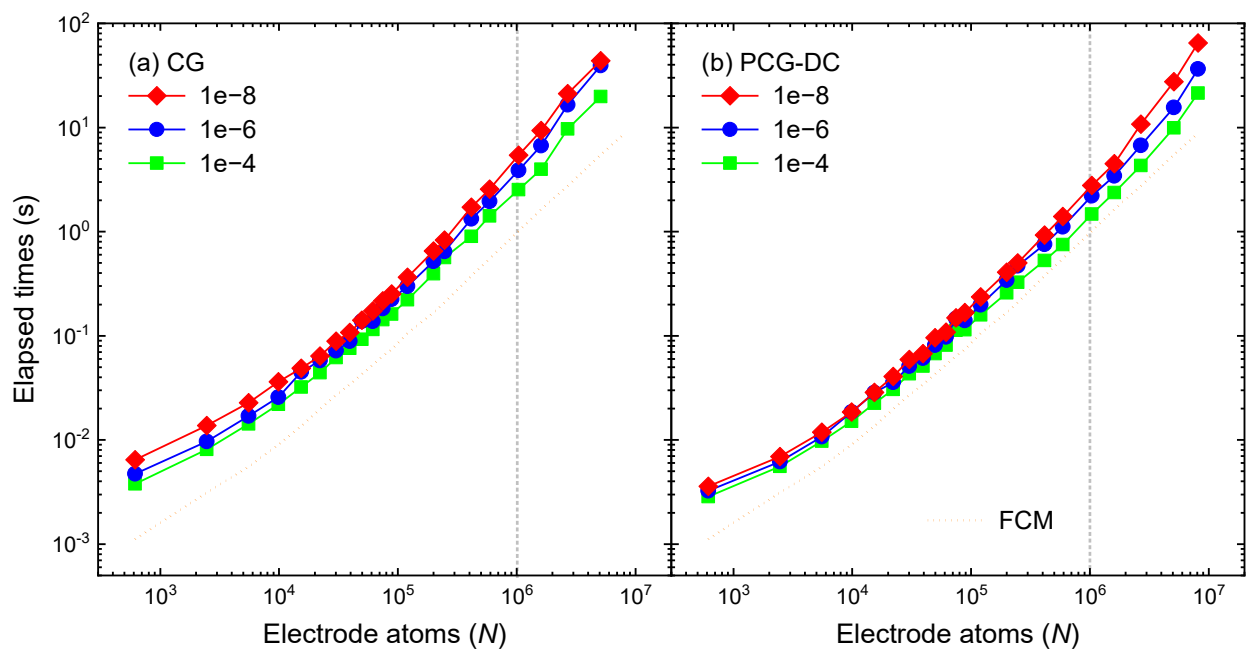

Figure S5: The time cost of an average MD step in the CG (a) or the PCG-DC (b) studies with the different convergent criterion, 1e−8 (red diamond), 1e−6 (blue circle) and 1e−4 (green square). The orange dashed line illustrates the average single-step MD cost by the conventional FCM method.

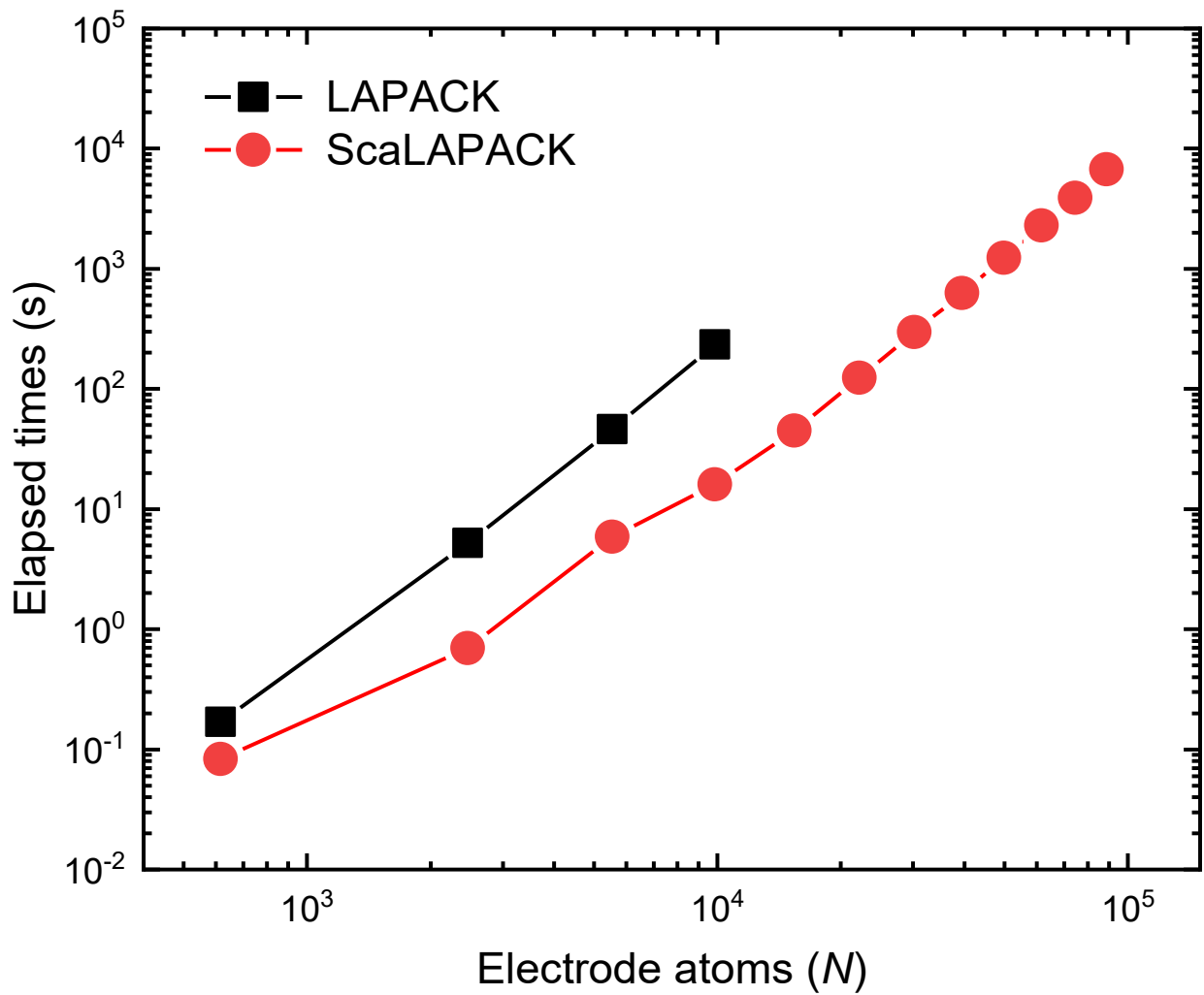

Figure S6: the time cost to inverse a dense Ewald interaction matrix using LAPACK (black square) and ScaLAPACK (red circle) libraries using 24 processors.

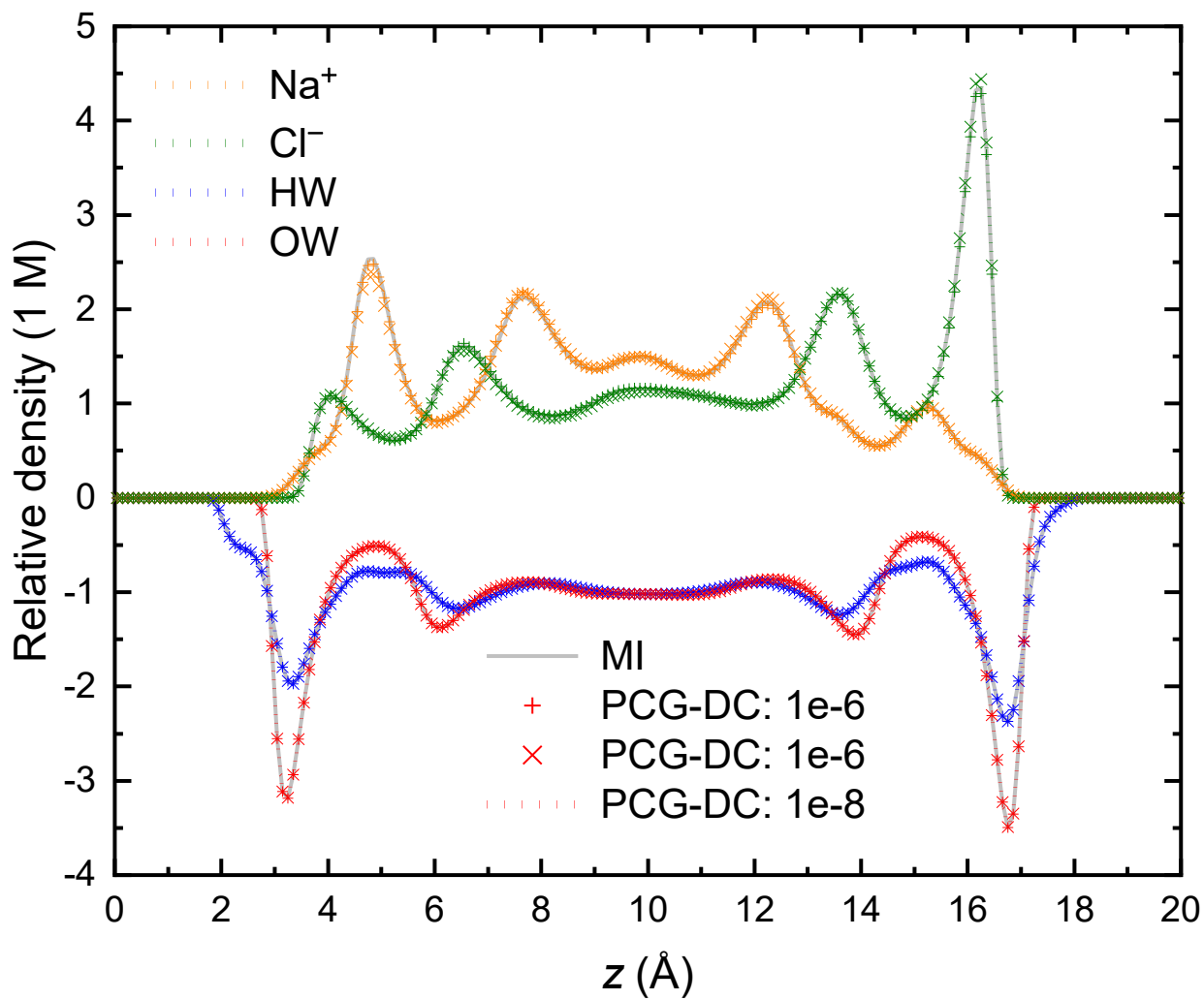

Figure S7: The EDLs along *z*-direction in 2-nm graphene electrochemical cell for PCG-DC studies under the different criteria, 1e−4 (+), 1e−6 (×) and 1e−8 (dotted), respectively. The grey curves illustrate the results via the MI approach.

**Reference**

[1] D. Borwein, J. Borwein, and R. Shail, J. Math. Anal. Appl. **143** (1989) 126.

[2] Z. Hu, J. Chem. Theory Comput. **10** (2014) 5254.

[3] D. Heyes, and A. Brańka, J. Chem. Phys. **138** (2013) 034504.

[4] G. Hummer, L. R. Pratt, and A. E. Garcia, J. Chem. Phys. **107** (1997) 9275.

[5] A. T. Celebi, S. H. Jamali, A. Bardow, T. J. Vlugt, and O. A. Moultos, Mol. Simul. **47** (2021) 831.

[6] S. Beane, W. Detmold, K. Orginos, and M. Savage, Prog. Part. Nucl. Phys. **66** (2011) 1.

[7] L. Zeng, T. Wu, T. Ye, T. Mo, R. Qiao, and G. Feng, (2021)

[8] J. C. Maxwell, *A Treatise on Electricity and Magnetism* (Cambridge University Press, 2010),

[9] J. Lekner, Journal of Electrostatics **69** (2011) 11.

[10] L. Scalfi, D. T. Limmer, A. Coretti, S. Bonella, P. A. Madden, M. Salanne, and B. Rotenberg, Phys. Chem. Chem. Phys. **22** (2020) 10480.